\documentclass{svjour3}                     
\smartqed  
\usepackage{graphicx}
\usepackage{amssymb, latexsym, textcomp,  textcomp}
\usepackage{amsmath}    
\journalname{JLTP}
\begin{document}

\title{Magneto-oscillations and anomalous current states in a photo-excited electron gas on liquid helium}

\titlerunning{Magneto-oscillations in a photo-excited electron gas on liquid helium}

\author{Yuriy Monarkha         \and         Denis Konstantinov}


\institute{Yu. Monarkha \at
              Institute for Low Temperature Physics and Engineering, 47 Nauky Avenue,
61103, Kharkiv, Ukraine. \\
              \email{monarkha@ilt.kharkov.ua}           
           \and
           D. Konstantinov \at
              Quantum Dynamics Unit, Okinawa Institute of Science and Technology (OIST) Graduate University, Tancha 1919-1, Okinawa 904-0495, Japan. \\
              \email{denis@oist.jp}           
}

\date{Received: date / Accepted: date}

\maketitle

\begin{abstract}
The paper reviews a novel class of phenomena observed recently in the
two-dimensional (2D) electron system formed on the free surface of liquid
helium in the presence of a magnetic field directed normally and exposed to
microwave radiation. The distinctive feature of these nonequilibrium
phenomena is magnetoconductivity oscillations induced by inter-subband
(out-of-plane) and intra-subband (in-plane) microwave excitations. The
conductivity magneto-oscillations induced by intra-subband excitation are
similar to remarkable microwave-induced resistance oscillations (MIRO)
reported for semiconductor heterostructures. Investigations of
microwave-induced conductivity oscillations (MICO) on liquid helium helped
with understanding of the origin of MIRO. Much stronger microwave-induced
conductivity oscillations were observed and well described theoretically for
resonant inter-subband microwave excitation. At strong powers, such
excitation leads to zero-resistance states (ZRS), the in-plane
redistribution of electrons, self-generated audio-frequency oscillations,
and incompressible states. These phenomena are caused by unusual current states of
the 2D electron system formed under resonant microwave excitation.
\end{abstract}

\section{Introduction}
\label{intro}
A two-dimensional (2D) electron gas on the free surface of liquid helium is
formed by a one-dimensional (1D) potential well $V\left( z\right) $ created by a
weak polarization attraction of an electron to the liquid medium and by a
strong repulsion barrier $V_{0}\simeq 1\,\mathrm{eV}$ at the
interface~\cite{ColCoh-1969,Shi-1970}. In an experiment with surface electrons (SE), a pressing
electric field $E_{\bot }$ is usually applied perpendicularly to the
surface, which gives a correction $eE_{\bot }z$ to $V\left( z\right) $. If
the source of electrons is not switched off, the equilibrium surface density
of electrons $n_{e}$ is proportional to the pressing field: $n_{e}=E_{\bot
}/2\pi e$. At this condition, the electric field of an electron layer $2\pi
en_{e}$ compensates the external field $E_{\bot }$ at large heights above
the layer: $z\gg 1/\sqrt{n_{e}}$. For the bulk liquid helium, electron
density is limited~\cite{LeiWan-1979,Ede-1980} by a critical value $n_{c}\simeq
2\cdot 10^{9}\,\mathrm{cm}^{-2}$. Employing helium films allows obtaining
substantially larger electron densities~\cite{EtzGomLei-1984}, still for
nonequilibrium phenomena considered in this review electron densities
usually are much smaller than $n_{c}$.

In the limit of weak fields ($E_{\bot }\rightarrow 0$), the energy spectrum
of an electron in the 1D potential well $\Delta _{l}$ is similar to the
spectrum of a hydrogen atom (the Rydberg states), $\Delta _{l}=-\Delta
/l^{2} $, where $l=1,2,...$ is the quantum number describing a surface
electron level. The characteristic energy $\Delta $ depends on the
dielectric constant of liquid helium $\epsilon $ as
\begin{equation}
\Delta =\frac{m_{e}\Lambda ^{2}}{2\hbar ^{2}}, \,\,\, \Lambda =\frac{%
e^{2}\left( \epsilon -1\right) }{4\left( \epsilon +1\right) },  \label{e1}
\end{equation}%
where $m_{e}$ and $-e$ are the free electron mass and charge respectively.
For liquid $^{4}\mathrm{He}$, $\Delta \simeq 7.64\,\mathrm{K}$, and the
average height $\left\langle 1\right\vert z\left\vert 1\right\rangle \simeq
114\,\mathrm{\mathring{A}}$. In the case of liquid $^{3}\mathrm{He}$, we
have $\Delta \simeq 4.3\,\mathrm{K}$ and $\left\langle 1\right\vert
z\left\vert 1\right\rangle \simeq 152\,\mathrm{\mathring{A}}$. The pressing
electric field $E_{\bot }$ increases the excitation energies $\Delta
_{l^{\prime },l}=\Delta _{l^{\prime }}-\Delta _{l}$ which allows to tune the
electron system to the resonance with the microwave (MW) field. By analogy
with semiconductor electron systems, electron states belonging to a certain
surface level ($l$) are usually called "surface subband".

Above the Wigner solid transition the in-plane states of SE are well
described by the 2D wave functions of free electrons and by the usual energy
spectrum. In the presence of a magnetic field directed perpendicular to the
interface, the in-plane electron states are squeezed into a set of Landau
levels: $\varepsilon _{n}=\hbar \omega _{c}\left( n+1/2\right) $, where $%
n=0,1,2,...$ , and $\omega _{c}=eB/cm_{e}$ is the cyclotron frequency. In
this case, surface electrons on liquid helium represent a singular system of
free particles with a purely discrete spectrum if interactions can be
neglected. Electron scattering by helium vapor atoms and capillary waves
(ripplons) leads to the collision broadening $\Gamma _{n}$ of the density of
states function which is usually described by a set of Gaussian functions.

Generally, the 2D electron system formed on liquid helium is similar to 2D
electron systems created in semiconductor devices~\cite{AndFowSte-1982} such as the
metal-oxide-semiconductor field-effect transistor or GaAs/AlGaAs heterostructures. The
broad study of the 2D electron gas in semiconductor devices revealed a
number of fundamental discoveries, of which the most known example is the
quantum Hall effect~\cite{KliDorPep-1980,TsuStoGos-82}. There are definite
differences between SE on helium and a 2D electron gas in semiconductors.
The most important one is that the effective mass of SE, which is very close to
the free electron mass $m_{e}$, is much larger than the effective mass of
electrons in semiconductor devices ($m_{e}\gg m_{e}^{\ast }$). This means
that, for actual electron densities, SE on liquid helium represent a
nondegenerate 2D electron gas with a strong Coulomb interaction between
electrons. Therefore, some well known quantum effects, such as
Shubnikov-deHaas oscillations, are impossible for SE on liquid helium.
Moreover, strong electron-electron correlations and the Wigner solid
transition~\cite{GriAda-1979} additionally suppress the quantum Hall
effect (for a review, see~\cite{MonSyv-2012}). On the other hand, SE levitate above a very clean surface
of liquid helium with no impurities and defects. The only scatterers
available are helium vapor atoms and capillary wave excitations (ripplons)
whose densities decrease with temperature. At typical helium temperatures ($%
T<0.5\,\mathrm{K}$), electrons have very high mobility and well defined Landau
levels with an extremely small collision broadening: $\Gamma _{n}\ll T$.
Therefore, SE on liquid helium represent a remarkable model system for
studying quantum magnetotransport phenomena~\cite{And-1997,MonTesWyd-2002,MonKon-book},
and this system is complementary to the 2D electron systems in semiconductor
devices.

At the beginning of this century, magnetotransport studies of a 2D electron
gas in high-mobility GaAs/AlGaAs heterostructures subjected to a dc magnetic
field and to strong microwave (MW) radiation led to an unexpected discovery:
MW-induced resistance oscillations (MIRO)~\cite{ZudDuRen-2001,YeEngRen-2001}.
The period of these oscillations is controlled only by the ratio of the microwave
frequency $\omega $ to the cyclotron frequency $\omega _{c}$, and, therefore, they
potentially can be expected in a nondegenerate 2D electron gas as well. At
high radiation power the minima of the oscillations evolve into
zero-resistance states (ZRS)~\cite{ManSmeKli-2002,ZudDuWes-2003}. Positions of the resistance
minima reported~\cite{ManSmeKli-2002} obey a universal law: $\omega /\omega
_{c}=m+1/4$ (here $m=1,2,...$). Especial interest in ZRS observed is
provoked by their plausible relationship with the concept of absolute
negative conductivity~\cite{AndAleMil-2003} associated with photon-assisted scattering of
electrons off impurities~\cite{Ryz-1969,DurSachRea-2003,RyzChaSur-2004,DmiMirPol-2003,DmiVavAle-2005}. These
discoveries have opened a new research area and triggered a large body of
theoretical works (for a review, see~\cite{DmiMirPol-2012}). The effect of
MIRO was observed also in hole systems~\cite{ZudMirEbn-2014}, and in MgZnO/ZnO
heterostructures~\cite{KarShcSme-2016}.

\begin{figure}
  \center{\includegraphics[width=8.5cm]{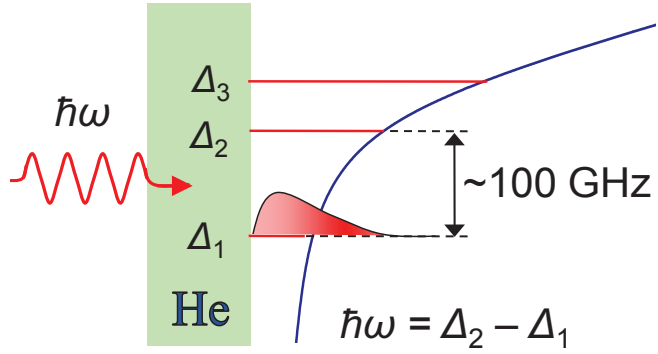}}
\caption{(Color online) Schematic view of inter-subband resonant excitation of SE on liquid helium}
\label{f1}
\end{figure}

\begin{figure}
 \center{ \includegraphics[width=10cm]{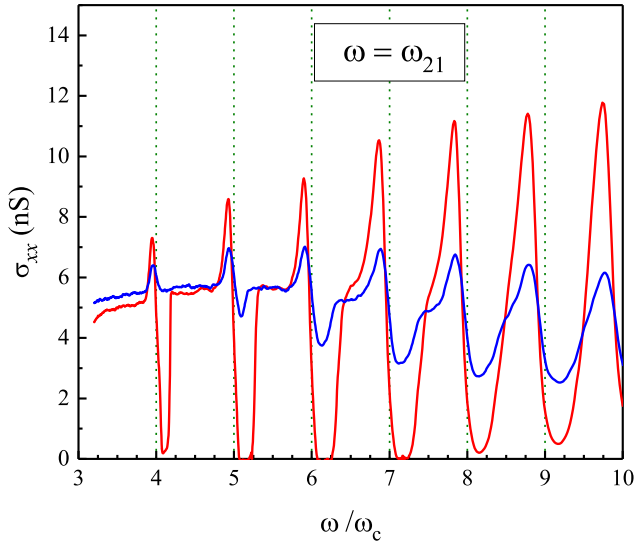}}
\caption{(Color online) Oscillations of $\sigma _{xx}$ induced by resonant MW excitation
obtained~\cite{KonKon-2010} for $T=0.2\,\mathrm{K}$, $n_{e}=0.9\cdot 10^{6}\,\mathrm{cm}^{-2}$, and two MW powers }
\label{f2}
\end{figure}

Similar MW-induced conductivity oscillations (MICO) and ZRS were discovered
in the 2D electron gas on liquid helium~\cite{KonKon-2009,KonKon-2010}.
For SE on liquid helium, we use the abbreviation MICO instead of MIRO because in experiments
the electron conductivity $\sigma _{xx}$ was actually measured using the Corbino setup.
The important difference of these oscillations, as compared to the MIRO in semiconductor
devices, is that they arise only when the excitation energy $\Delta
_{2,1}=\Delta _{2}-\Delta _{1}\equiv \hbar \omega _{2,1}$ is tuned to the
resonance with the MW field ($\Delta _{2,1}=\hbar \omega $) using the linear
Stark effect for the 1D potential well $V\left( z\right) $ as illustrated in
Fig.~\ref{f1}. Typical magnetoconductivity ($\sigma _{xx}$) oscillations
of SE induced by the resonant MW
excitation are shown in Fig.~\ref{f2} for two values of excitation power.
The absence of MICO for $\omega $ values slightly different from the
condition $\hbar \omega =\Delta _{2,1}$ means that the period of these
oscillations is actually controlled by the ratio $\Delta _{2,1}/\hbar \omega
_{c}$ and the inter-subband (out-of-plane) excitation of SE is crucial for
understanding the origin of these oscillations. Since the mechanisms
proposed for explanation of the MIRO in semiconductors assume a pure
intra-subband MW excitation, the MICO reported for SE on liquid
helium~\cite{KonKon-2009,KonKon-2010} must have different origin.
This conclusion is confirmed also by a
noticeable increase of the amplitude of MICO with the ratio $\omega /\omega
_{c}$ at $m\leq 7$, which is opposite to the observation reported for
semiconductor heterostructures. The explanation of remarkable features of
MICO observed on liquid helium were given by the theory~\cite{Mon-2011,Mon-2012}
based on a nonequilibrium population of the first
excited surface subband which leads to sign-changing terms in the electron
magnetoconductivity and even to absolute negative conductivity ($\sigma
_{xx}<0$) at high radiation power.

Experimental investigations of MICO induced by inter-subband MW excitation
revealed a number of new phenomena: the resonant photovoltaic effect and
spatial redistribution of electrons~\cite{KonCheKon-2012}, self-generated
audio-frequency oscillations~\cite{KonWatKon-2013}, and an incompressible
state~\cite{CheWatKon-2015}. The all these effects occur near the experimental condition
$\sigma _{xx}\rightarrow 0$ which allows assuming that the instability of
the spatially uniform distribution of SE is caused by absolute negative
conductivity. Moreover, the self-generation of audio-frequency current
oscillations at an electrode indicates that initially the damping of these
oscillations is negative which also agrees with the concept of negative
conductivity. A plausible explanation of these novel observations is based
on the Coulombic effect on the stability range of the photo-excited electron
gas which favors formation of domains of different densities~\cite{Mon-2016}.
Domains appear to eliminate or greatly reduce regions with negative
conductivity.

The mechanisms of MIRO based on photon-assisted scattering of
electrons~\cite{Ryz-1969,DurSachRea-2003,DmiMirPol-2003} and reported for
a degenerate 2D electron gas in
semiconductors can also be applied to SE on liquid helium~\cite{Mon-2014}.
In this case, the probability of the intra-subband photon-assisted
scattering is proportional to $E_{\mathrm{mw}}^{2}/m_{e}^{2}$, where $E_{%
\mathrm{mw}}$ is the amplitude of the MW field. Therefore, in order to
obtain the same amplitude of MIRO, the microwave field should be $%
m_{e}/m_{e}^{\ast }$ times larger than in semiconductor systems. Taking into
account that $m_{e}^{\ast }\simeq 0.06\,m_{e}$, it was estimated that $E_{%
\mathrm{mw}}$ should be about $10\,\mathrm{V/cm}$ to induce a substantial
amplitude of oscillations. In a recent experiment~\cite{YamMonKon-2015},
employing a semiconfocal Fabry-Perot resonator the in-plane electric filed $%
E_{\mathrm{mw}}$ of necessary amplitude was created in the layer of SE, and
MW-induced dc magnetoconductivity oscillations governed by the ratio $%
\omega /\omega _{c}$ (analogous to MIRO in GaAs/AlGaAs heterostructures)
were observed. This observation proved the universality of the effect of MIRO,
and gave rise to hopes that additional investigations of MICO in the system
of SE on helium will help with identification of its origin. By present
time, these hopes were justified at least partly by a discovery of a strong
dependence of MICO on the MW circular polarization direction~\cite{ZadMonKon-2018}.
This discovery allowed reporting the first observation of
the effect of radiation helicity, which provides crucial information for
understanding the origin of intra-subband MICO in a 2D electron gas. In particular, these
experiments unambiguously support theoretical mechanisms of MICO based on
photon-assisted scattering off disorder.

The advances in experimental and theoretical investigations of
nonequilibrium phenomena in the 2D electron gas on liquid helium induced by
MW radiation have motivated us to write this review.

\section{Mechanisms of MICO and negative conductivity}
\label{sec:2}

As noted above, for SE on liquid helium, MICO was initially observed for resonant
inter-subband excitation as indicated in Fig.~\ref{f1}. Nevertheless, it is instructive to begin
discussions of theoretical mechanisms of MICO with intra-subband models, assuming that the all SE occupy
the ground surface subband.

\subsection{Intra-subband models}
\label{subsec:2-1}
It is very surprising that by now there is a great body of different
theoretical mechanisms explaining MIRO in semiconductor devices which use
quantum and classical effects (see the review~\cite{DmiMirPol-2012}), but the origin of
these oscillations is still under debate. Among these theories, there is a
large group of models whose description is based on the concept of the
photon-assisted scattering off disorder which overcomes the selection rules
existing for direct photon-induced transitions (direct transitions can be
only between adjacent Landau levels).

\subsubsection{Displacement model}
\label{subsubsec:2-1-1}
Magneto-oscillations and absolute negative conductivity induced by
MW radiation in thin semiconductor films in transverse magnetic fields were
predicted by Ryzhii~\cite{Ryz-1969} already in 1969. The physics of this
effect is quite simple. In the presence of a driving dc electric field $%
\mathbf{E}_{\Vert }$ directed along the $x$-axis, the Landau spectrum
acquires a correction depending on the center coordinate of the cyclotron
motion ($X$):
\begin{equation}
\varepsilon _{n,X}=\hbar \omega _{c}\left( n+1/2\right) +eE_{\Vert }X%
.  \label{e2}
\end{equation}%
For an elastic scattering off disorder $n,X\rightarrow n^{\prime },X^{\prime
}$ accompanied by absorption of a photon, the energy conservation yields%
\begin{equation}
\hbar \omega _{c}\left( n^{\prime }-n\right) +eE_{\Vert }\left( X^{\prime
}-X\right) -\hbar \omega =0.  \label{e3}
\end{equation}%
Using the notation $n^{\prime }-n=m>0$, one can represent this equation in the following form%
\begin{equation}
X^{\prime }-X=\frac{\hbar \omega _{c}}{eE_{\Vert }}\left( \frac{\omega }{%
\omega _{c}}-m\right) .  \label{e4}
\end{equation}%
This simple relationship indicates that $X^{\prime }-X$ (the direction of
scattering) changes its sign when the ratio $\omega /\omega _{c}$ passes an
integer $m=1,2,...$ . For example, at $\omega /\omega _{c}>m$, the difference
$X^{\prime }-X>0$ which means that an electron is scattered in the direction
opposite to the direction of the driving force $-e\mathbf{E}_{\Vert }$.
Actually, this is a qualitative explanation of the origin of absolute negative conductivity.

The relationship of Eq.~(\ref{e4}) is just a form of the energy conservation
in a scattering event which is not convenient for the analysis of the limiting case $E_{\Vert }\rightarrow 0$.
Nevertheless, the factor $(\omega /\omega _{c} -m)$ entering its right side
remarkably appears in an accurate conductivity treatment.
In the magnetoconductivity treatment discussed below, one have to take into account also
the momentum conservation restricting $X'-X$ and the collision broadening of Landau
levels restricting the factor $(\omega /\omega _{c} -m)$ and selecting the number $m$
responsible for the major contribution to $\sigma _{xx}$ at a given magnetic field.
More specifically, the left side of Eq.~(\ref{e3}) represents the argument of the delta-function
describing the probability of scattering, and the later is expanded in $eE_{\Vert }\left( X^{\prime
}-X\right)$ to obtain the linear conductivity $\sigma _{xx}$.

The Eq.~(\ref{e4}) explains why this mechanism of MIRO
is called "displacement model". For semiconductor heterostructures, it
was developed in many works~\cite{DurSachRea-2003,RyzChaSur-2004,VavAle-2004} (more
references can be found in the review~\cite{DmiMirPol-2012}). For SE on liquid
helium, this model was extended~\cite{Mon-2014,Mon-2017} to include strong
Coulomb interaction between electrons. Considering the MW electric field as
a classical field $\mathbf{E}_{\mathrm{mw}}\left( t\right) $, it is possible
to show~\cite{Mon-2017} that the contribution from photon emission processes
contains an additional exponential factor $\exp \left( -\hbar \omega
/T\right) $ which allows neglecting these processes at low temperatures.

Equations describing MICO can be found in a quite simple way by
introducing the in-plane momentum exchange $\hbar \mathbf{q}$ at a
collision. Indeed, $\sigma _{xx}$ can be calculated using clear
relationships for the current%
\begin{equation}
j_{x}=-en_{e}\sum_{\mathbf{q}}\left( X^{\prime }-X\right) _{\mathbf{q}}\bar{w%
}_{\mathbf{q}}\left( V_{\mathrm{H}}\right) =-en_{e}\ell _{B}^{2}\sum_{%
\mathbf{q}}q_{y}\bar{w}_{\mathbf{q}}\left( V_{\mathrm{H}}\right) ,
\label{e5}
\end{equation}%
where $\ell _{B}=\left( \hbar c/eB\right) ^{1/2}$ is the magnetic length, $%
\bar{w}_{\mathbf{q}}\left( V_{\mathrm{H}}\right) $ is the average
probability of electron scattering with the momentum exchange $\hbar \mathbf{%
q}$ which is a function of the Hall velocity $V_{\mathrm{H}}=cE_{\Vert }/B$.
In Eq.~(\ref{e5}), we have used the usual momentum conservation equation $%
X^{\prime }-X=q_{y}\ell _{B}^{2}$ which follows from matrix elements of the
operator $e^{-i\mathbf{q}\cdot \mathbf{r}}$ calculated for the Landau states
(free electron states $\mathbf{k}^{\prime}$ and $\mathbf{k}$ obviously give
$\mathbf{k}^{\prime}-\mathbf{k}=\mathbf{q}$). This equation defines also the dependence $\bar{w}%
_{\mathbf{q}}\left( V_{\mathrm{H}}\right) $, because the energy conservation
delta-function contains $eE_{\Vert }\left( X^{\prime }-X\right) =\hbar
q_{y}V_{\mathrm{H}}$ (it is instructive to note that the energy exchange $\hbar q_{y}V_{\mathrm{H}}$
appears also for quasi-elastic scattering in the moving reference frame~\cite{MonKon-book} where $E^{\prime }_{\Vert } =0$).
The Eq.~(\ref{e5}) yields
\begin{equation}
\sigma _{xx}\simeq \frac{e^{2}n_{e}\nu _{\mathrm{eff}}}{m_{e}\omega _{c}^{2}}%
,\,\,\,\,\,\, \nu _{\mathrm{eff}}=-\frac{1}{m_{e}V_{\mathrm{H}}}\sum_{\mathbf{q}%
}\hbar q_{y}\bar{w}_{\mathbf{q}}\left( V_{\mathrm{H}}\right) .  \label{e6}
\end{equation}%
Expanding $\bar{w}_{\mathbf{q}}\left( V_{\mathrm{H}}\right) $ in $V_{\mathrm{%
H}}$ one can obtain the linear dc magnetoconductivity. Usually this
procedure leads to the additional $q_{y}=\left( X^{\prime }-X\right) /\ell
_{B}^{2}$ in the integrand of Eq.~(\ref{e6}), and we have $q_{y}^{2}$
independent of the sign of $X^{\prime }-X$. Therefore, the appearance of
sign-changing terms in the expressions for $\nu _{\mathrm{eff}}$
and $\sigma _{xx}$ is not that trivial as it might be concluded from Eq.~(\ref{e4}).
Actually, these terms appear because the probability of scattering
$\bar{w}_{\mathbf{q}}$ as a function of the energy exchange has maxima at Landau
excitation energies $m\hbar \omega _{c}$, and the derivative of $\bar{w}_{\mathbf{q}}$ near the
maxima yields the necessary terms.

The most difficult part of the description of the displacement model is to
obtain $\bar{w}_{\mathbf{q}}\left( V_{\mathrm{H}}\right) $ for the
photon-assisted scattering. The Hamiltonian of the electron--ripplon
interaction, which dominates at low temperatures, can be written in the form
similar to that of the electron-phonon interaction%
\begin{equation}
V_{\mathrm{e-r}}=\frac{1}{\sqrt{S_{A}}}\sum_{\mathbf{q}}U_{q}Q_{q}\left( b_{-%
\mathbf{q}}+b_{\mathbf{q}}^{\dag }\right) e^{-i\mathbf{q}\cdot \mathbf{r}},
\label{e7}
\end{equation}%
where $S_{A}$ is the surface area (in the following, it will be
set to unity), $\mathbf{r}$ is a 2D in-plane
radius-vector of an electron, $b_{\mathbf{q}}^{\dag }$ and $b_{\mathbf{q}}$
are creation and distraction operators of ripplons, $U_{q}$ is the
electron-ripplon coupling~\cite{ShiMon-74}, $Q_{q}^{2}=\hbar
q/2\rho \omega _{r,q}$, $\omega _{r,q}\simeq \sqrt{\alpha /\rho }q^{3/2}$ is
the ripplon spectrum, $\alpha $ and $\rho $ are the surface tension and mass
density of liquid helium respectively. The electron-ripplon coupling
contains the contribution from polarization interaction with liquid helium
and the pressing field term ($eE_{\bot }$).

There are two ways of finding scattering probabilities for the
photon-assisted scattering. In the pure quantum approach, the
electron-photon interaction $V_{\mathrm{e-p}}$ is proportional to the vector
potential of the MW field $\mathbf{A}_{\mathrm{mw}}$ expressed in terms of
creation and destruction operators of photons. Then, probabilities of the
photon-assisted scattering are calculated~\cite{Mon-2014}
according to the Golden Rule with matrix elements $\left\langle f\right\vert
\tilde{V}\left\vert i\right\rangle $ containing products of matrix elements
of $V_{\mathrm{e-r}}$ and $V_{\mathrm{e-p}}$. Still, there is a more elegant
and nonperturbative way using the Landau-Floquet states (for
examples, see Refs.~\cite{Mon-2017,Par-2004}; the method is based on the
early work of Husimi~\cite{Hus-1953}). In this case, the interaction
with the MW field is described using the classical correction to the
Hamiltonian: $eE_{\mathrm{mw}}^{\left( x\right) }\left( t\right) x+eE_{%
\mathrm{mw}}^{\left( y\right) }\left( t\right) y$. Then, using the Landau
gauge for the vector potential of the magnetic field $B$, one can find
the Landau-Floquet eigenfunctions $\psi _{n,X}^{\left( \mathrm{F}\right)
}\propto \varphi _{n}\left( x-X-\xi ,t\right) $ containing some time
dependent parameters [like $\xi \left( t\right) $] and factors chosen to
reduce $\varphi _{n}\left( x,t\right) $ to the usual oscillator
eigenfunction. The important point is that $n$ and $X$ remain good quantum
numbers so that we can use Eqs.~(\ref{e5}) and (\ref{e6}) for obtaining $%
\sigma _{xx}$. Still, the additional time-dependent parameters and factors
of $\psi _{n,X}^{\left( \mathrm{F}\right) }$ affect the matrix elements of $%
V_{\mathrm{e-r}}$ and the energy conservation delta-function.

It is useful to specify the dependence $\mathbf{E}_{\mathrm{mw}}\left(
t\right) $ as%
\begin{equation}
E_{\mathrm{mw}}^{\left( x\right) }=aE_{0}\cos \omega t, \,\, E_{%
\mathrm{mw}}^{\left( y\right) }=bE_{0}\sin \omega t,  \label{e8}
\end{equation}%
where $E_{0}$ is the amplitude parameter of the MW field, and, generally, $a$
and $b$ are arbitrary parameters. For particular cases, we assume that $a$
can be $0$ or $1,$ and $b$ can be $0$ or $\pm 1$. Thus, we can describe two
linear polarizations [parallel ($a=1$, $b=0$) and perpendicular ($a=0$, $b=1$%
) to the dc electric field] and two circular polarizations ($a=1$, $b=\pm 1$%
). Respectively, we define the polarization index $p=\Vert ,\bot ,+,-$ ,
where the first two symbols ($\Vert $ and $\bot $) correspond to linear
polarizations, and the last two symbols ($+$ and $-$) correspond to circular
polarizations.

Scattering probabilities depend on the matrix elements $\left( e^{-i\mathbf{q%
}\cdot \mathbf{r}}\right) _{n^{\prime },X^{\prime },n,X}^{\left( \mathrm{F}%
\right) }$ calculated for the Landau-Floquet states which have an additional
factor
\begin{equation}
\exp \left[ -i\beta _{p,\mathbf{q}}\sin \left( \omega t+\gamma \right) %
\right] \label{e9}
\end{equation}
as compared to the usual matrix elements $\delta _{X,X^{\prime }-\ell
_{B}^{2}q_{y}}\left( e^{-iq_{x}x}\right) _{n^{\prime },X^{\prime
};n,X}^{\left( \mathrm{L}\right) }$ obtained in the absence of the MW field.
Here the parameter $\beta _{p,\mathbf{q}}$ depends strongly on the MW
polarization
\begin{equation}
\beta _{p,\mathbf{q}}=\frac{\lambda \omega _{c}\ell _{B}}{\left( \omega
^{2}-\omega _{c}^{2}\right) }\sqrt{q_{y}^{2}\left( a\omega _{c}+b\omega
\right) ^{2}+q_{x}^{2}\left( a\omega +b\omega _{c}\right) ^{2}},  \label{e10}
\end{equation}%
$\lambda =eE_{0}l_{B}/\hbar \omega $ describes the strength of the MW field,
and the phase shift $\gamma $ is unimportant for the following. It should be
emphasized that, in the presence of MW radiation, we still have the momentum
conservation rule $X^{\prime }-X=q_{y}\ell _{B}^{2}$ used in Eq.~(\ref{e5}).
By means of the Jacobi-Anger expansion $e^{iz\sin \phi }=\sum_{k}J_{k}\left(
z\right) e^{ik\phi }$ [here $J_{k}\left( z\right) $ is the Bessel function]
one can find that scattering probabilities are proportional to the sum of
delta-functions~\cite{Mon-2017}%
\[
w_{\mathbf{q};n,X\rightarrow n^{\prime },X^{\prime }}^{\left( \pm \right) }=%
\frac{2\pi }{\hbar }\left\vert C_{\mathbf{q}}^{\left( \pm \right)
}\right\vert ^{2}I_{n,n^{\prime }}^{2}\left( x_{q}\right) \delta
_{X,X^{\prime }-l_{B}^{2}q_{y}}\times
\]%
\begin{equation}
\times \sum_{k=-\infty }^{\infty }J_{k}^{2}\left( \beta _{p,\mathbf{q}%
}\right) \delta \left( \varepsilon _{n^{\prime },X^{\prime }}-\varepsilon
_{n,X}-k\hbar \omega \pm \hbar \omega _{r,q}\right) ,  \label{e11}
\end{equation}%
representing the processes of absorption or emission of $\left\vert
k\right\vert $ photons along with absorption or emission of a ripplon. Here $%
x_{q}=q^{2}\ell _{B}^{2}/2$ is a dimensionless variable, $C_{\mathbf{q}%
}^{\left( \pm \right) }=\left( U_{q}\right) _{1,1}Q_{q}\left[ n_{\pm \mathbf{%
q}}^{(r)}+1/2\pm 1/2\right] ^{1/2}$, $n_{\mathbf{q}}^{(r)}$ is the
number of ripplons with the wave vector $\mathbf{q}$,
\begin{equation}
I_{n,n^{\prime }}^{2}\left( x_{q}\right) =\frac{\min \left( n,n^{\prime
}\right) !}{\max \left( n,n^{\prime }\right) !}x_{q}^{\left\vert n^{\prime
}-n\right\vert }e^{-x_{q}}\left[ L_{\min \left( n,n^{\prime }\right)
}^{\left\vert n^{\prime }-n\right\vert }\left( x_{q}\right) \right] ^{2},
\label{e12}
\end{equation}%
and $L_{n}^{m}\left( x_{q}\right) $ are the associated Laguerre polynomials.
Since the ripplon energy $\hbar \omega _{r,q}$ is usually much smaller than
typical electron energies, it can be neglected in the argument of the
delta-functions. It is remarkable that considering the MW field in a pure
classical way we found that the energy exchange between the field and an
electron equals to an integer number of quanta of the electromagnetic
field ($k\hbar \omega $).

According the relationship $X^{\prime }-X=q_{y}\ell _{B}^{2}$, the quantity
$\sum_{X^{\prime }}w_{\mathbf{q};n,X\rightarrow n^{\prime },X^{\prime }}^{\left( \pm \right) }$
to be used for obtaining $\bar{w}_{\mathbf{q}}\left( V_{\mathrm{H}}\right) $
of nondegenerate SE is independent of $X$, and, therefore, it can be averaged over discrete
Landau numbers $n$ only using the distribution function $e^{-\varepsilon _{n}/T_{e}}/Z_{\parallel }$,
where $Z_{\parallel }$ is the partition function for the spectrum
$\varepsilon _{n}=\hbar \omega _{c}\left( n+1/2\right) $.
For a 2D electron gas under magnetic field, one have to take into account the collision
broadening of Landau levels to obtain a finite result. In the self-consistent Born approximation (SCBA)~\cite{AndUem-1974},
level densities have a semi-elliptic shape. The cumulant approach~\cite{Ger-1976} yields a Gaussian
shape with the same broadening parameter $\Gamma _{n}$. Generally, the level shape is a kind of
average of elliptical and Gaussian forms~\cite{And-1974}, and the lowest level is shaped like a Gaussian.
In the following we shall use the Gaussian form because it simplifies evaluations and it is preferable
for low levels. Introducing Landau level densities of states~\cite{Ger-1976}%
\begin{equation}
g_{n}\left( \varepsilon \right) =\frac{\sqrt{2\pi }\hbar }{\Gamma _{n}}\exp %
\left[ -\frac{2\left( \varepsilon -\varepsilon _{n}\right) ^{2}}{\Gamma
_{n}^{2}}\right] ,  \label{e13}
\end{equation}%
with a finite collision broadening $\Gamma _{n}$, the average probability of
electron scattering $\bar{w}_{\mathbf{q}}\left( V_{\mathrm{H}}\right) $ with
the momentum exchange $\mathbf{q}$, can be found in terms of the dynamic
structure factor (DSF) $S\left( q,\Omega \right) $ of a nondegenerate 2D
electron gas~\cite{Mon-2017}
\begin{equation}
\bar{w}_{\mathbf{q}}\left( V_{\mathrm{H}}\right) =2u_{1,1}^{2}\sum_{k=-\infty }^{\infty }J_{k}^{2}\left(
\beta _{s,\mathbf{q}}\right) S\left( q,k\omega -q_{y}V_{\mathrm{H}}\right) .
\label{e14}
\end{equation}%
Here we introduce $u_{l,l^{\prime }}^{2}=\left\vert \left( U_{q}\right)
_{l,l^{\prime }}\right\vert ^{2}Q_{q}^{2}N_{r,q}/\hbar ^{2}$ defined by matrix
elements of the electron-ripplon coupling $U_{q}\left( z\right) $, and the
ripplon distribution function $N_{r,q}$. The DSF of the 2D Coulomb liquid
under a magnetic field can be represented as a sum of Gaussians~\cite{MonKon-book}
\begin{equation}
S\left( q,\Omega \right) \simeq \frac{2\sqrt{\pi }}{Z_{\parallel }}%
\sum_{n,n^{\prime }}e^{-\varepsilon _{n}/T_{e}}\frac{I_{n,n^{\prime }}^{2}}{%
\gamma _{n;n^{\prime }}}
\exp \left[ -\frac{\left[ \Omega -\left( n^{\prime }-n\right) \omega
_{c}-\phi _{n}\right] ^{2}}{\gamma _{n;n^{\prime }}^{2}}\right] .
\label{e15}
\end{equation}%
It is quite usual that peaks of the DSF as a function of
frequency occur at excitation energies of the system, while interactions
affect the broadening of these peaks. If the Coulomb interaction can be
neglected, the broadening of the Gaussians is determined by the average
broadening of the respective Landau levels $\Gamma _{n;n^{\prime }}=\sqrt{%
\Gamma _{n}^{2}/2+\Gamma _{n^{\prime }}^{2}/2}$. The Coulomb interaction
affects the parameters $\gamma _{n;n^{\prime }}$ and $\phi _{n}$ in the
following way~\cite{MonKon-book}
\begin{equation}
\hbar \gamma _{n;n^{\prime }}=\sqrt{\Gamma _{n;n^{\prime }}^{2}+x_{q}\Gamma
_{C}^{2}},\,\,\, \phi _{n}=\frac{\Gamma _{n}^{2}+x_{q}\Gamma
_{C}^{2}}{4T_{e}\hbar },  \label{e16}
\end{equation}%
where $\Gamma _{C}=\sqrt{2}eE_{\mathrm{f}}^{(0)}\ell _{B}$ and $E_{\mathrm{f}%
}^{\left( 0\right) }\simeq 3\sqrt{T_{e}}n_{e}^{3/4}$ is the typical value of
the internal electric field of the fluctuational origin~\cite{FanDykLea-1997}.

Above given equations for the DSF of the Coulomb liquid were
obtained~\cite{MonTesWyd-2002,MonKon-book} considering
an ensemble of noninteracting electrons whose orbit centers are moving fast in the
uniform fluctuational electric field $\mathbf{E}_{\mathrm{f}}$. Therefore, plasmon
excitations are not present in this DSF.
It is remarkable that Eqs.~(\ref{e15}) and (\ref{e16}) describing the DSF of
SE are valid even for the Wigner solid state~\cite{MonKon-book} if $\omega _{c}$
is much larger than the typical frequency of longitudinal phonons.
At low electron densities $n_{e}$, the shift $\phi _{n}$ can be neglected
because $\Gamma _{n}\ll T$ under usual experimental conditions. Thus, $%
S\left( q,\Omega \right) $ has sharp maxima at Landau excitation frequencies
$\Omega \simeq m\omega _{c}$ which appeared to be a very useful property for
the description of MICO.

The main features of the displacement model can be easily seen from the
expression%
\begin{equation}
\nu _{\mathrm{eff}}=\frac{2\hbar }{m_{e}}\sum_{\mathbf{q}%
}q_{y}^{2}u_{1,1}^{2}\sum_{k=-\infty }^{\infty }J_{k}^{2}\left( \beta _{p,%
\mathbf{q}}\right) S^{\prime }\left( q,k\omega \right)   \label{e17}
\end{equation}%
which is a direct consequence of Eqs.~(\ref{e6}) and (\ref{e14}), if $\bar{w}%
_{\mathbf{q}}\left( V_{\mathrm{H}}\right) $ is expanded up to the linear
term in $q_{y}V_{\mathrm{H}}$. The important point is that $\nu _{\mathrm{eff}}$ contains the
derivative of the DSF: $S^{\prime }\left( q,k\omega \right) $. The presence
of the derivative of the function $S\left( q,\Omega \right) $ which has
sharp maxima at Landau excitation frequencies ($m\omega _{c}$, $m=1,2...$)
explains the appearance of sign-changing terms, in spite of the positive factor
$q_{y}^{2}$. At the same time, for usual scattering processes (without any
photon involved; $k=0$), the basic property of the equilibrium DSF
$S\left( q,-\Omega \right) =\exp \left( -\hbar \Omega /T_{e}\right) S\left( q,\Omega
\right) $ gives $S^{\prime }\left( q,0\right) =\left(
\hbar /2T_{e}\right) S\left( q,0\right) >0$ and, therefore,
Eq.~(\ref{e17}) reproduces the result of the
SCBA theory~\cite{AndFowSte-1982,MonKon-book,Ger-1976}
if $\Gamma _{C}\rightarrow 0$.
The same property of the DSF allows transforming $\nu _{\mathrm{eff}%
}=\sum_{k=0}^{\infty }\nu _{k}$ where the terms with $k\geq 1$ have an
additional factor $\left( 1-e^{-k\hbar \omega /T_{e}}\right) $ (here some
small corrections were neglected). The second term of this factor represents
the contribution from photon emission processes; it can be neglected
at low temperatures.

The term with $k=1$ describes one-photon-assisted scattering. According to
Eq.~(\ref{e17}), $\nu _{1}$ contains the sum of derivatives of the Gaussian
functions entering $S\left( q,\omega \right) $:
\begin{equation}
\nu _{1}\propto -\sum_{m}\frac{2\left( \omega -m\omega _{c}-\phi _{n}\right)
}{\gamma _{n;n^{\prime }}^{2}}\exp \left[ -\frac{\left( \omega -m\omega
_{c}-\phi _{n}\right) ^{2}}{\gamma _{n;n^{\prime }}^{2}}\right] ,
\label{e18}
\end{equation}%
where $m=n^{\prime }-n$. From this equation one can see that $\nu _{1}<0$
when $\omega /\omega _{c}>m$ which is in accordance with Eq.~(\ref{e4}).
Thus, Eqs.~(\ref{e17}) and (\ref{e18}) describe the shape of MIRO and MICO
which agrees with expectations based on the qualitative analysis of Eqs.~(\ref%
{e3}) and (\ref{e4}) and with experimental observations. For small broadening
of the DSF maxima, the exponential factor of Eq.~(\ref{e18}) selects the
number $m\equiv n^{\prime }-n=1,2,...$ which gives the major contribution
into $\nu _{1}$ for a chosen $\omega $. This factor is close to unity only
if $\omega -m\omega _{c}\simeq 0$, and it is exponentially small for other $%
m $ or $\omega _{c}$ which do not meet this condition.

\begin{figure}
 \center{ \includegraphics[width=10cm]{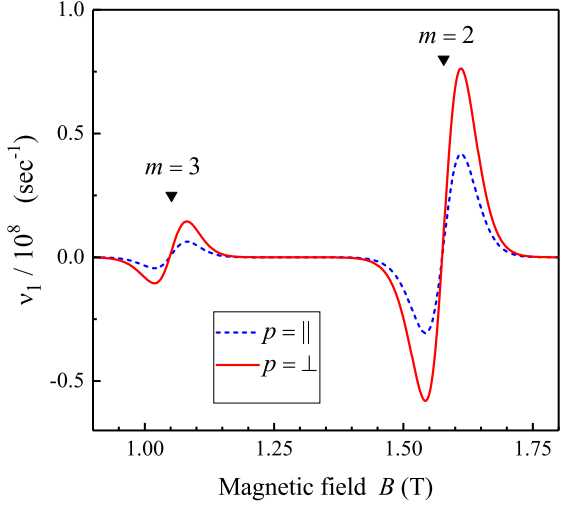}}
\caption{(Color online) The oscillatory contribution to the effective collision frequency
$\nu _{1}$ vs. $B$ calculated for two different MW polarizations:
parallel (to the dc electric field) $p=\Vert$ (blue dashed) and perpendicular $p=\bot$ (red solid)~\cite{Mon-2017}.
Black triangles indicate the values of $B$ such that $\omega /\omega _c=m$.
The conditions are the following: $E_{0}=10\,\mathrm{V/cm}$, $n_{e}=17\cdot 10^{6}\,\mathrm{cm}^{-2}$,
$\omega /2\pi =88.34\,\mathrm{GHz}$, and $T=0.56\,\mathrm{K}$ (liquid $^{4}\mathrm{He}$) }
\label{f3}
\end{figure}

The effective collision frequency induced by one-photon-assisted scattering
$\nu _{1}$ is shown in Fig.~\ref{f3} for two linear MW
polarizations. Firstly, we note that the minima occur at lower fields $B$ than the
respective maxima. Secondly, the amplitude of intra-subband MICO decreases
steady and strongly with $m$. It is interesting to note also that in the
limit of large $\gamma _{n;n^{\prime }}$ (strong collision broadening $%
\Gamma _{n}$), the overlapping of the sign-changing terms in the sum over
the all $m$ limits the position of minima by a universal law $\omega /%
\omega _{c}\rightarrow m+1/4$, which coincides remarkably with that
reported in Ref.~\cite{ManSmeKli-2002}.

\subsubsection{Inelastic model}
\label{subsubsec:2-1-2}

When averaging $w_{\mathbf{q};n,X\rightarrow n^{\prime },X^{\prime
}}^{\left( \pm \right) }$ given in Eq.~(\ref{e11}) we naturally used the
equilibrium electron distribution function $f\left( \varepsilon \right) $, which is
indicated in Eq.~(\ref{e15}) by the Boltzmann factor $e^{-\varepsilon
_{n}/T_{e}}$.\ For the terms with $k>0$, in the regime $\beta _{p,\mathbf{q}%
}\ll 1$, this is a correct procedure. Still, as proven in Ref.~\cite{DmiMirPol-2003}%
, the photon-assisted scattering
leads to an oscillatory correction to the distribution function $f\left(
\varepsilon \right) $. This correction cannot be neglected in Eq.~(\ref{e17})
if we consider the term with $k=0$, because its contribution to $\nu _{%
\mathrm{eff}}=\sum_{k=0}^{\infty }\nu _{k}$ can be comparable with (or even
larger than) $\nu _{1}$ given by the displacement model.

In order to obtain the oscillating correction to $f\left( \varepsilon \right) $,
we restrict ourselves to scattering events involving only one photon
[$k=\pm 1$ in Eq.~(\ref{e11})] and assume $\beta _{p,\mathbf{q}}\ll 1$.
Analyzing the average transition rate up ($n\rightarrow n^{\prime }$) and
the all transition rates down ($n^{\prime }\rightarrow n$), we can represent
them as integral forms ($\int d\varepsilon \int d\varepsilon ^{\prime }...$)
using the Landau level density of states $g_{n}\left( \varepsilon \right) $
in the way that was used for finding the DSF $S\left( q,\Omega \right) $.
Then, it is possible to obtain the rate-balance condition~\cite{Mon-2017}%
\begin{equation}
f_{n^{\prime }}\left( \varepsilon ^{\prime }\right) =\frac{%
f_{n}\left( \varepsilon ^{\prime }-\hbar \omega \right) r_{n,n^{\prime
}}\left( \varepsilon ^{\prime }\right) }{\nu _{n^{\prime }}^{\left( \mathrm{%
2R}\right) }+r_{n,n^{\prime }}\left( \varepsilon ^{\prime }\right) },
\label{e19}
\end{equation}%
where
\begin{equation}
r_{n,n^{\prime }}\left( \varepsilon ^{\prime }\right) =\frac{\lambda ^{2}T}{%
2\hbar }\bar{\chi}_{p}\nu _{R}P_{n,n^{\prime }}g_{n}\left( \varepsilon
^{\prime }-\hbar \omega \right)   \label{e20}
\end{equation}%
is the excitation rate, $\nu _{R}=\Lambda ^{2}/8\pi \hbar \alpha \ell
_{B}^{4}$ is a characteristic collision frequency, the dimensionless
parameter $P_{n,n^{\prime }}$ describes the strength of the electron-ripplon
coupling in the presence of a magnetic field
\begin{equation}
P_{n,n^{\prime }}=\frac{\ell _{B}^{4}}{\Lambda ^{2}}\int\limits_{0}^{\infty
}U_{q}^{2}I_{n,n^{\prime }}^{2}\left( x_{q}\right) dx_{q},  \label{e21}
\end{equation}%
$\bar{\chi}_{p}$ is the polarization factor
\begin{equation}
\bar{\chi}_{\bot }=\bar{\chi}_{\Vert }=\frac{2\omega _{c}^{2}\left( \omega
_{c}^{2}+\omega ^{2}\right) }{\left( \omega ^{2}-\omega _{c}^{2}\right) ^{2}}%
,\,\,\bar{\chi}_{\pm }=\frac{4\omega _{c}^{2}\left( \omega \pm \omega
_{c}\right) ^{2}}{\left( \omega ^{2}-\omega _{c}^{2}\right) ^{2}},
\label{e22}
\end{equation}%
$\nu _{n^{\prime }}^{\left( \mathrm{2R}\right) }$ is the inelastic
transition rate from $n^{\prime }$ to the all $n<n^{\prime }$ caused by
two-ripplon emission processes~\cite{MonSokStu-2010}. In the distribution
function $f_{n^{\prime }}\left( \varepsilon ^{\prime }\right) $, the
subscript $n^{\prime }$ indicates that its argument $\varepsilon ^{\prime }$
is close to $\varepsilon _{n^{\prime }}$.

The Eq.~(\ref{e19}) reminds the solution of the rate equation of a two-level
model usually obtained in quantum optics. The second term in the denominator
of this equation is caused by backward electron transitions accompanied by
emission of a photon. Here the inelastic decay rate $\nu _{n^{\prime
}}^{\left( \mathrm{2R}\right) }$ plays an important role in obtaining the
oscillatory correction to the distribution function. Firstly, we note that
in the absence of $\nu _{n^{\prime }}^{\left( \mathrm{2R}\right) }$, the
solution of Eq.~(\ref{e19}) satisfies the saturation condition $f_{n^{\prime
}}\left( \varepsilon ^{\prime }\right) =f_{n}\left( \varepsilon ^{\prime
}-\hbar \omega \right) $ which is quite obvious. Secondly, a sharp shape of $%
f_{n^{\prime }}\left( \varepsilon ^{\prime }\right) $ appears when the
inelastic scattering rate is stronger then the excitation rate: $\nu
_{n^{\prime }}^{\left( \mathrm{2R}\right) }\gg r_{n,n^{\prime }}$. In this
limiting case, one can neglect $r_{n,n^{\prime }}$ in the denominator of the
right side of Eq.~(\ref{e19}) and set $f_{n}\left( \varepsilon \right) $ to
the equilibrium function in the numerator. This treatment yields $%
f_{n^{\prime }}\left( \varepsilon ^{\prime }\right) \propto g_{n}\left(
\varepsilon ^{\prime }-\hbar \omega \right) $, which means that at higher
Landau levels we have a sort of population inversion eventually leading to
MICO. This is the reason why this mechanism is called "the inelastic model".

If electron-electron interactions are neglected, the inelastic model gives
an additional correction to the effective collision frequency~\cite{Mon-2017}%
\[
\nu _{\mathrm{in}}=\pi \lambda ^{2}\bar{\chi}_{p}\nu _{R}^{2}T^{2}\hbar
\omega _{c}\sum_{n^{\prime }=1}^{\infty }\frac{P_{n^{\prime },n^{\prime
}}P_{0,n^{\prime }}}{\nu _{n^{\prime }}^{\left( 2R\right) }\Gamma
_{n^{\prime }}G_{0;n^{\prime }}}\times
\]
\begin{equation}
\times \frac{2\left( n^{\prime }\hbar \omega _{c}-\hbar \omega \right) }{%
G_{0;n^{\prime }}^{2}}\exp \left[ -\frac{\left( n^{\prime }\hbar \omega
_{c}-\hbar \omega \right) ^{2}}{G_{0;n^{\prime }}^{2}}\right] ,  \label{e23}
\end{equation}%
where $G_{n;n^{\prime }}^{2}=\Gamma _{n;n^{\prime }}^{2}-\Gamma _{n^{\prime
}}^{2}/4$. The Eq.~(\ref{e23}) indicates that, in the inelastic model, the
shape of MICO represents the derivative of a Gaussian similar to that of the
displacement model [see Eq.~(\ref{e18})]. In contrast with the displacement
model, additional large parameters $\nu _{R}/\nu _{n^{\prime }}^{\left(
2R\right) }$ and $T/G_{0;n}$ appear in the expression for $\nu _{\mathrm{in}}
$ which can make MICO more pronounced.

\begin{figure}
 \center{ \includegraphics[width=10cm]{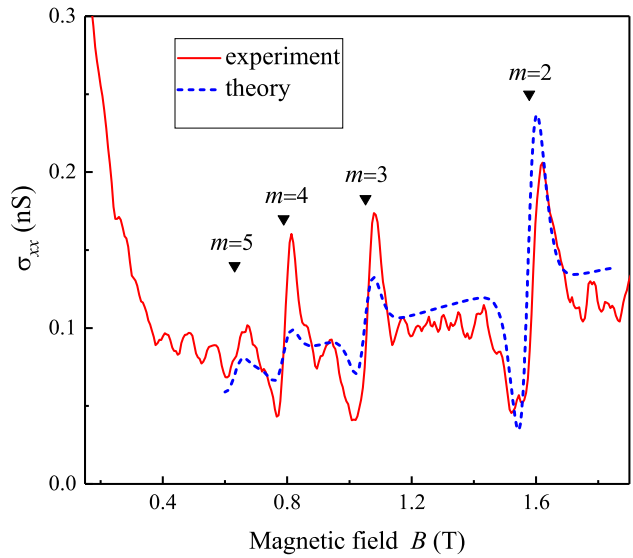}}
\caption{(Color online) $\sigma _{xx}$ versus $B$ for SE on liquid $^{4}\mathrm{He}$.
Data (red) and theoretical results (blue dashed) were obtained for
$n_{e}=1.7\cdot 10^{7}\,\mathrm{cm}^{-2}$, $T=0.56\,\mathrm{K}$ and $\omega /2\pi =88.34\,\mathrm{GHz}$~\cite{YamMonKon-2015}
 }
\label{f4}
\end{figure}

Unfortunately, in the inelastic model, it is very difficult to describe the
effect of Coulomb interaction on magneto-conductivity oscillations. Anyway,
it is reasonable to expect that the fluctuational electric field $\mathbf{E}%
_{f}$ will increase the broadening of oscillations in the way similar to
that of the displacement model. Typical conductivity variations caused by
the inelastic model are shown in Fig.~\ref{f4} by the blue dashed
line. The experimental data~\cite{YamMonKon-2015} shown here by the red
curve were obtained at a rather high MW frequency, where the contribution
from the displacement model is small.

\subsection{ Inter-subband model}
\label{subsec:2-2}

For MICO induced by inter-subband MW excitation of SE on liquid
helium~\cite{KonKon-2009,KonKon-2010}, contrary to MIRO in heterostructures,
there is only one theoretical mechanism proposed by now~\cite{Mon-2011,Mon-2012}.
Generally, it gives a quite well description of experimental observations.
It is remarkable that this inter-subband model cannot be reduced to any
intra-subband model of MIRO; moreover, it is not relevant to the
photon-assisted scattering which plays the major role in the displacement
and inelastic models. Nevertheless, it has something in common with both
the displacement and inelastic models. This can be easily seen even from
a qualitative analysis similar to that given above in Eqs.~(\ref{e3}) and (\ref{e4}).

\begin{figure}
 \center{ \includegraphics[width=8cm]{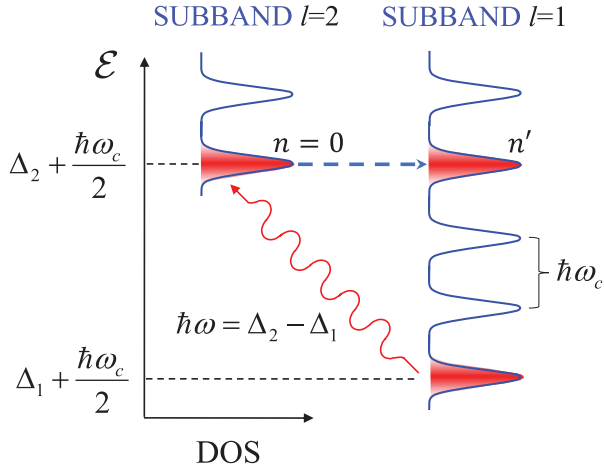}}
\caption{(Color online) Dynamics of SE in perpendicular
magnetic fields. MW photons of energy $\hbar \omega $ drive the
transition $l=1\rightarrow 2$ (wavy arrow) without changing the quantum
state $n$. Under the level matching condition, excited electrons can be scattered elastically (dashed blue
arrow) and fill the state $n^{\prime } >n$ of the ground subband~\cite{KonKon-2010}
 }
\label{f5}
\end{figure}

Consider an inter-subband scattering event $l,n,X\rightarrow l^{\prime
},n^{\prime },X^{\prime }$. Now we have to include the Rydberg state energy $%
\Delta _{l}$ into the electron spectrum in the presence of the magnetic and
driving electric fields
\begin{equation}
\mathcal{E}_{l,n,X}=\Delta _{l}+\hbar \omega _{c}\left( n+1/2\right)
+eE_{\Vert }X .  \label{e24}
\end{equation}%
For usual elastic scattering off disorder (photons are not involved in this
process), the energy conservation yields%
\begin{equation}
\hbar \omega _{c}\left( n^{\prime }-n\right) +eE_{\Vert }\left( X^{\prime
}-X\right) -\Delta _{l,l^{\prime }}=0 ,  \label{e25}
\end{equation}%
where we used the obvious relationship $\Delta _{l^{\prime },l}=\Delta
_{l^{\prime }}-\Delta _{l}=-\Delta _{l,l^{\prime }}$. For electron decay
processes from the first excited subband ($l=2$) to the ground subband ($%
l^{\prime }=1$), this scattering event is possible only if the magnetic
field is close to the level matching condition $\hbar \omega _{c}\left(
n^{\prime }-n\right) \approx \Delta _{2,1}$, as shown in Fig.~\ref{f5},
because $X^{\prime }-X$ is
limited by the magnetic length $\ell _{B}$ ($q\sim 1/\ell _{B}$). Under
other conditions, quasi-elastic decay caused by a ripplon is impossible.

Assuming a decay process from the first excited subband down
to the ground subband, and introducing the inter-subband excitation
frequency $\omega _{2,1}=\Delta _{2,1}/\hbar >0$, from Eq.~(\ref{e25}) one
can find the displacement of the electron orbit center
\begin{equation}
X^{\prime }-X=\frac{\hbar \omega _{c}}{eE_{\Vert }}\left( \frac{\omega _{2,1}%
}{\omega _{c}}-m\right) .  \label{e26}
\end{equation}%
Here $n^{\prime }-n=m>0$ because in this process an electron scatters to a
higher Landau level as illustrated in Fig.~\ref{f5}.
Eq.~(\ref{e26}) is very similar to Eq.~(\ref{e4}) used
above for explaining MIRO and the effect of absolute negative conductivity
in the displacement model. The only difference is that now we have the inter-subband
excitation frequency $\omega _{2,1}$ instead of the MW frequency $%
\omega $.

At this stage, photons are not necessary to cause electron scattering
against the driving force $-e\mathbf{E}_{\Vert }$. For example,
quasi-elastic inter-subband scattering from $l=2$ to $l^{\prime}=1$ will be the
scattering against the driving force ($X^{\prime }>X$), if $\omega
_{2,1}/\omega _{c}>m$, or when $B$ is a bit lower than the level matching
point $\omega _{2,1}/\omega _{c}=m$. The important thing is that there are
inverse scattering processes from $l=1$ to $l^{\prime }=2$. For these processes,
the right side of the respective equation for $X^{\prime }-X$ \ [similar to
Eq.~(\ref{e26})] has the opposite sign (minus) with $m=n-n^{\prime }>0$,
because an electron scatters to a lower Landau level.
Therefore, at the same conditions $\omega _{2,1}/\omega _{c}>m$,
electron scattering up the surface subbands is obviously the scattering
along the driving force ($X^{\prime }<X$). The average probability of
scattering $\bar{\nu}_{l\rightarrow l^{\prime }}$ of SE
usually satisfies the condition $\nu _{l^{\prime }\rightarrow l}=\nu
_{l\rightarrow l^{\prime }}\exp \left( -\hbar \omega _{l,l^{\prime
}}/T_{e}\right) $, where  $\hbar \omega _{l,l^{\prime }}=\Delta _{l}-\Delta _{l^{\prime }}$.
Thus, we can expect that the sign-changing correction to $\sigma _{xx}$
induced by inter-subband scattering will have the following form
\begin{equation}
\delta \sigma _{xx}^{\left( \mathrm{inter}\right) }\propto -\left(
N_{2}-N_{1}e^{-\hbar \omega _{2,1}/T}\right) \sum_{m}\left( \frac{\omega
_{2,1}}{\omega _{c}}-m\right) e^{-\left( \omega _{2,1} -m\omega _{c}\right)
^{2}/\gamma _{n;n^{\prime }}^{2}},  \label{e27}
\end{equation}%
where $N_{l}$ is the number of electrons at the level $l$. In the right side
of this equation, the exponential proportionality factor is introduced in
order to select the number $m$ giving the major contribution to $\delta
\sigma _{xx}^{\left( \mathrm{inter}\right) }$ similar to Eq.~(\ref{e18}). At
equilibrium, we obviously have $N_{2}/N_{1}=e^{-\hbar \omega _{2,1}/T}$ and,
therefore, $\delta \sigma _{xx}^{\left( \mathrm{inter}\right) }=0$. To
obtain the sign-changing corrections to $\sigma _{xx}$ and even the absolute
negative conductivity we have to create an extra population of the first
excited level
\begin{equation}
N_{2}>N_{1}e^{-\hbar \omega _{2,1}/T},  \label{e28}
\end{equation}%
which can be naturally induced by the resonant MW excitation with $\omega =$
$\omega _{2,1}$ shown in Fig.~\ref{f5} by the wavy arrow.

In Eq.(\ref{e27}), the factor $\left( \omega _{2,1}/\omega _{c}-m\right) $,
describing qualitatively the inter-subband mechanism of MICO, reminds the
factor $\left( \omega /\omega _{c}-m\right) $ of the displacement model of MIRO.
At the same time, the first factor $\left(
N_{2}-N_{1}e^{-\hbar \omega _{2,1}/T}\right) $ requires a nonequilibrium
electron distribution over surface subbands which has something in common
with the inelastic model. Nevertheless, in contrast with the inelastic model,
here the condition of Eq.~(\ref{e28}) is created by direct resonant
absorption of a MW quantum (without involving any kind of disorder), and the
population inversion (important for the inelastic model) is not necessary.
Moreover, both the displacement and inelastic models require photon-assisted
scattering as the origin of MIRO, while the inter-subband mechanism
remarkably have no relation to the photon-assisted scattering described in
Subsection~\ref{subsec:2-1}. In the inter-subband model, photons are used only for
providing a nonequilibrium population of
the excited subband; the sign-changing correction to $\sigma _{xx}$ and
absolute negative conductivity are caused by usual scattering off disorder in
a nonequilibrium multi-subband 2D electron system. In the qualitative analysis
given above, we discussed only sign-changing corrections to $\sigma _{xx}^{\left( \mathrm{inter}\right) }$.
The accurate treatment presented below indicates that there is also a normal (remaining positive) contribution
to $\sigma _{xx}^{\left( \mathrm{inter}\right) }$ which exists even for
$N_{2}/N_{1}=e^{-\hbar \omega _{2,1}/T}$, but it is less important if
the MW power is strong enough.

The magneto-conductivity treatment given in Eqs.~(\ref{e5}) and (\ref{e6})
can be extended to include inter-subband scattering
\begin{equation}
\nu _{\mathrm{eff}}=-\frac{1}{m_{e}V_{\mathrm{H}}}\sum_{l,l^{\prime }}\bar{n}%
_{l}\sum_{\mathbf{q}}\hbar q_{y}\bar{w}_{l,l^{\prime }}\left( \mathbf{q},V_{%
\mathrm{H}}\right) ,  \label{e29}
\end{equation}%
where $\bar{n}_{l}=N_{l}/N_{e}$ are the fractional occupancies of surface
subbands, and $\bar{w}_{l,l^{\prime }}\left( \mathbf{q},V_{\mathrm{H}%
}\right) $ is the average probability of both intra and inter-subband
scattering ($l\rightarrow l^{\prime }$) which is accompanied by the momentum
exchange $\hbar \mathbf{q}$. For usual scattering processes
(MW photons are not involved in a scattering event), $\bar{w}%
_{l,l^{\prime }}\left( \mathbf{q},V_{\mathrm{H}}\right) $ is found
as~\cite{Mon-2013_LTP,Mon-2013_JETP}
\begin{equation}
\bar{w}_{l,l^{\prime }}\left( \mathbf{q}\right) =2u_{l,l^{\prime
}}^{2}\left( x_{q}\right) S_{l,l^{\prime }}\left( q,\omega _{l,l^{\prime
}}-q_{y}V_{\mathrm{H}}\right) ,  \label{e30}
\end{equation}%
where the function $u_{l,l^{\prime }}^{2}\left( x_{q}\right) $ was
introduced just under Eq.~(\ref{e14}) and $S_{l,l^{\prime }}\left( q,\Omega
\right) $ is an extension of the DSF $S\left( q,\Omega \right) $ given in
Eq.~(\ref{e15}) applicable for a multi-subband 2D electron system, because
now we have to take into account that the broadening of a Landau level
depends also on $l$ ($\Gamma _{n}\rightarrow \Gamma _{l,n}$). For intra-subband
scattering ($l=l^{\prime }$), the right side of Eq.~(\ref{e30}) transforms
naturally into the term with the photon number $k=0$ of Eq.~(\ref{e14}).
According to Eq.~(\ref{e30}), the case $\omega _{l,l^{\prime }}>0$ ($\omega _{l,l^{\prime }}<0$) resembles
electron scattering accompanied by absorbtion (emission) of a photon whose
frequency $\omega =\left\vert \omega _{l,l^{\prime }}\right\vert $.

The accurate form of $S_{l,l^{\prime }}\left( q,\Omega \right) $ can be formally obtained
from the definitions of $S\left( q,\Omega \right) $ given in Eqs.~(\ref{e15})
and (\ref{e16}) using the following replacements: $\Gamma _{n}\rightarrow
\Gamma _{l,n}$, $\Gamma _{n;n^{\prime }}\rightarrow \Gamma _{l,n;l^{\prime
},n^{\prime }}$, $\gamma _{n;n^{\prime }}\rightarrow \gamma _{l,n;l^{\prime
},n^{\prime }}$, and $\phi _{n}\rightarrow \phi _{l,n}$. We assume that
electron distribution over Landau levels can be still described by the
Boltzmann function with an effective temperature $T_{e}$ due to the strong
Coulomb interaction. In this case, $S_{l,l^{\prime }}\left( q,\Omega \right)
$ has an important property%
\begin{equation}
S_{l,l^{\prime }}\left( q,-\Omega \right) =e^{-\hbar \Omega
/T_{e}}S_{l^{\prime },l}\left( q,\Omega \right) ,  \label{e31}
\end{equation}%
which simplifies conductivity evaluations. The effective temperature approximation used
here is based on the experimental fact~\cite{ZipBroGri-1976} that the electron-velocity
autocorrelation time $\tau _{c}$ is usually much shorter than all other relaxation times.
For example, at a low electron density $n_{e}=1.5\cdot 10^{7}\,\mathrm{cm}^{-2}$, the reciprocal
value $\tau _{c}^{-1}\simeq 10^{10}\,\mathrm{sec}^{-1}$ is close to the harmonic
oscillator frequency in a 2D triangular lattice $\omega_{0} $, which means that the electron
system resembles the Wigner solid. Even for the smallest density used in MICO experiments
$n_{e}\simeq 10^{6}\,\mathrm{cm}^{-2}$, the average Coulomb interaction energy of SE is much
larger than the average kinetic energy which means that the energy exchange between electrons is
very strong. It should be emphasized additionally that the expression for the DSF of the Coulomb liquid given
above remarkably coincides with the DSF of the Wigner solid~\cite{MonKon-book} heated to $T_{e}$.

\begin{figure}
 \center{ \includegraphics[width=10cm]{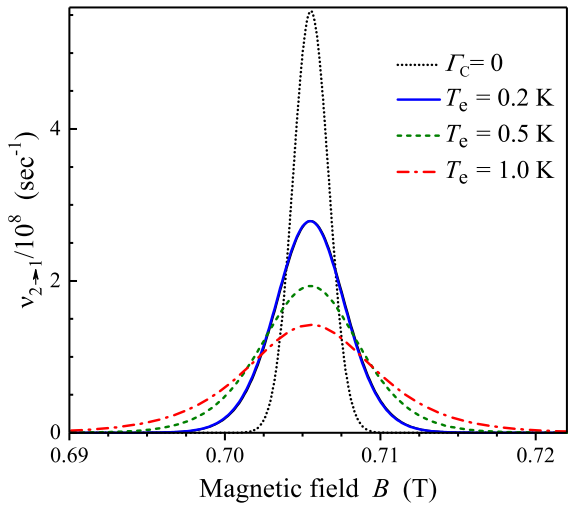}}
\caption{(Color online) Decay rate of the first excited subband $\nu _{2\rightarrow 1}$ vs.
the magnetic field $B$ for $T=0.2\,\mathrm{K}$ and $n_{e}=10^{6}\,\mathrm{cm}^{-2}$:
single-electron treatment (dotted); many-electron theory was calculated for different $T_{e}$
as indicated in the figure legend~\cite{Mon-2012}
 }
\label{f6}
\end{figure}

The effective collision frequency of Eq.~(\ref{e29}) depends on the
fractional occupancies $\bar{n}_{l}$ which should be found from the rate
equation (similar to that of the quantum optics) which contains the decay
rate of the excited subband $\nu _{l\rightarrow l^{\prime }}$. Using Eq.~(\ref%
{e30}), the later quantity can be found as%
\begin{equation}
\nu _{l\rightarrow l^{\prime }}=2\sum_{\mathbf{q}}u_{l,l^{\prime
}}^{2}\left( x_{q}\right) S_{l,l^{\prime }}\left( q,\omega _{l,l^{\prime
}}\right) .  \label{e32}
\end{equation}%
In this equation, $q_{y}V_{\mathrm{H}}$ is set to zero because we consider
the linear dc transport properties. Thus, the decay rate $\nu _{2\rightarrow
1}$ has sharp maxima near the level matching points: $\omega _{2,1}=m\omega
_{c}$. The typical dependence of the decay rate $\nu _{2\rightarrow 1}$ on
the magnetic field near the level matching point is shown in Fig.~\ref{f6}.
The strong temperature dependence of $\nu _{2\rightarrow 1}$
maxima is caused by the Coulomb correction $\Gamma _{C}$
to $\gamma _{l,n;l^{\prime },n^{\prime }}$. The above noted Coulomb
broadening of the decay rate $\nu _{2\rightarrow 1}$ affects strongly
subband occupancies $\bar{n}_{l}=N_{l}/N$ in the presence of MW radiation.
For the two-subband model, which is valid for $T_{e}\leq 2\,\mathrm{K}$,
the conventional rate equation yields~\cite{KonMonKon-2007}
\begin{equation}
\frac{\bar{n}_{2}}{\bar{n}_{1}}=\frac{1+\exp \left( -\hbar \omega
_{2,1}/T_{e}\right) \nu _{2\rightarrow 1}/r_{\mathrm{mw}}}{1+\nu
_{2\rightarrow 1}/r_{\mathrm{mw}}},  \label{e33}
\end{equation}%
where $r_{\mathrm{mw}}$ is the MW excitation rate. Under the resonance
condition, $r_{\mathrm{mw}}=\Omega _{\mathrm{R}}^{2}/2\gamma _{\mathrm{mw}}$%
, where $\gamma _{\mathrm{mw}}$ is the half-width of the resonance, and $%
\Omega _{\mathrm{R}}$ is the Rabi frequency proportional to the amplitude of
the MW field. According to Eq.~(\ref{e33}) and Fig.~\ref{f6}, the ratio $\bar{n%
}_{2}/\bar{n}_{1}$ oscillates with $B$ having minima near the level matching
points and approaching the saturation condition $\bar{n}_{2}/\bar{n}%
_{1}\rightarrow 1$ between these points. Usually, even a small nonequilibrium
filling of the excited subband $\bar{n}_{2}-\bar{n}_{1}e^{-\hbar \omega _{2,1}/T_{e}}>0$
can lead to giant oscillations in $\sigma_{xx}$ due to $\Gamma _{l,n}\ll T_{e}$.

The Eqs.~(\ref{e29}), (\ref{e32}) and (\ref{e33}) allow describing MICO
induced by nonequilibrium population of the first excited subband. For example,
consider scattering only between $l=2$ and $l=1$ (the two-subband model).
Using the property of the DSF
given in Eq.~(\ref{e31}), the contribution of inter-subband scattering $\nu _{%
\mathrm{inter}}$ to the effective collision frequency $\nu _{\mathrm{eff}}$
can be represented as a sum of two distinctive terms $\nu _{\mathrm{inter}%
}=\nu _{\mathrm{A}}+\nu _{\mathrm{N}}$, where
\begin{equation}
\nu _{\mathrm{A}}=\frac{2\hbar }{m_{e}}\left( \bar{n}_{2}-\bar{n}%
_{1}e^{-\hbar \omega _{2,1}/T_{e}}\right) \sum_{\mathbf{q}%
}q_{y}^{2}u_{2,1}^{2}S_{2,1}^{\prime }\left( q,\omega _{2,1}\right) ,
\label{e34}
\end{equation}%
\begin{equation}
\nu _{\mathrm{N}}=\frac{2\hbar ^{2}}{m_{e}T_{e}}\bar{n}_{1}e^{-\hbar \omega
_{2,1}/T_{e}}\sum_{\mathbf{q}}q_{y}^{2}u_{2,1}^{2}S_{2,1}\left( q,\omega
_{2,1}\right) .  \label{e35}
\end{equation}%
The first term $\nu _{\mathrm{A}}$ represents an anomalous (sign-changing)
contribution which is proportional to the derivative of the sum of Gaussians
$S_{2,1}^{\prime }\left( q,\omega _{2,1}\right) $. Obviously, the shape of
conductivity variations near the level matching points originated from this
term is similar to that of MIRO and MICO caused by photon-assisted
scattering. It is important that at equilibrium ($\bar{n}_{2}=\bar{n}%
_{1}e^{-\hbar \omega _{2,1}/T_{e}}$), this term vanishes. The second term $%
\nu _{\mathrm{N}}$ represents a normal contribution which oscillates with $B$
remaining positive. One can use slightly different definitions~\cite{Mon-2012}
of $\nu _{\mathrm{A}}$ and $\nu _{\mathrm{N}}$ by subtracting
\begin{equation}
\frac{\hbar ^{2}}{m_{e}T_{e}}\left( \bar{n}_{2}-\bar{n}_{1}e^{-\hbar \omega
_{2,1}/T_{e}}\right) \sum_{\mathbf{q}}q_{y}^{2}u_{2,1}S_{2,1}\left( q,\omega
_{2,1}\right)  \label{e36}
\end{equation}%
from the right side of Eq.~(\ref{e34}) and adding it into Eq.~(\ref{e35}).
Such redistribution does not change $\nu _{\mathrm{inter}}$
but leads to a more symmetrical form of $\nu _{\mathrm{N}}$
proportional to $\bar{n}_{2}+\bar{n}_{1}e^{-\hbar \omega _{2,1}/T_{e}}$.
Anyway, for low electron densities, $\hbar \gamma _{l,n;l^{\prime
},n^{\prime }}\ll T$ and, therefore, the corrections of Eqs.~(\ref{e35}) and (%
\ref{e36}) are much smaller that $\nu _{\mathrm{A}}$ defined in Eq.~(\ref{e34}%
).

\begin{figure}
 \center{ \includegraphics[width=10.5cm]{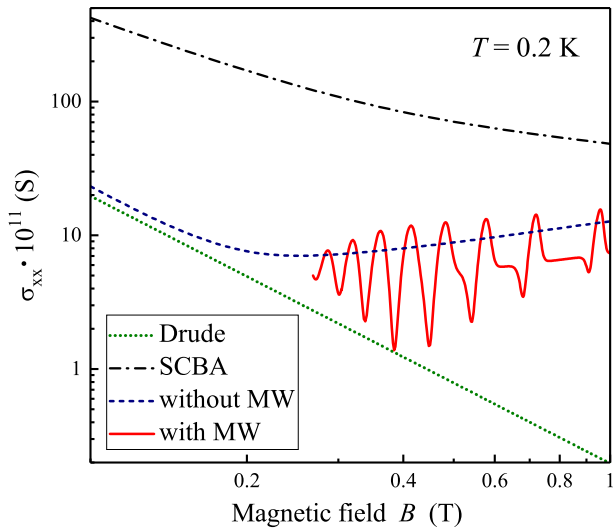}}
\caption{(Color online) Magnetoconductivity vs. $B$ calculated for $T=0.2\,\mathrm{K}$ ($^{3}\mathrm{He}$) and
$n_{e}=10^{7}\,\mathrm{cm}^{-2}$: the SCBA theory (dash-dotted),
the Drude approximation (olive dotted), the dark many-electron theory (blue dashed), the theory based on the
inter-subband mechanism of MICO (red solid)~\cite{Mon-2012}
 }
\label{f7}
\end{figure}

Typical MICO calculated for MW radiation of medium power are shown in
Fig.~\ref{f7} by the solid red line. Here the single electron theory (dark) based on the SCBA
is shown by the dash-dotted line. Without MW excitation the many electron theory
(blue dashed) transforms from the
SCBA result to the Drude approximation when $B$ decreases.
It should be noted that the amplitude of inter-subband MICO firstly increases with lowering $B$ (increasing $m$), but then,
at lower $B$, it decreases and vanishes due to the Coulombic effect.
Eventually $\sigma _{xx}$ is approaching the many-electron line (blue
dashed) calculated for $\mathbf{E}_{\mathrm{mw}}=0.$ This behavior is in
accordance with experimental observations~\cite{KonKon-2009,KonKon-2010}.

The multi-subband electron system on liquid helium can be tuned in resonance with
the MW field for electron transitions to higher subbands ($l>2$). In this
case, the dc magnetic ($B$) and electric ($E_{\bot }$) fields oriented normally to the
multisubband 2D electron system give a remarkable possibility to manipulate
the inter-subband scattering probabilities and to realize the population
inversion of electron subbands~\cite{Mon-2013_JETP}. The theoretical analysis
of the electron momentum relaxation rate under the resonant MW excitation of
the third subband ($l=3$) indicates that such an excitation induces a
variety of new magneto-oscillations of $\sigma _{xx}$. Among these, there
are oscillations, accompanying by the population inversion ($\bar{n}_{2}>%
\bar{n}_{1}$), with a period which is incommensurate with the basic period
determined by the resonant MW frequency, oscillations with a $1/B$ periodic
amplitude modulation, and oscillations located in the vicinity of some
fractional values of the ratio $\omega /\omega _{c}$.

\section{The Coulombic effect on MICO}
\label{sec:3}

In order to test theoretical mechanisms and models by an experiment it is
always good to have variable parameters affecting the outcome. For SE on
liquid helium, one of the important parameters is the electron density $%
n_{e} $ which defines the strength of the Coulomb interaction between
electrons. In a nondegenerate electron system, the average Coulomb
interaction energy per an electron $U_{\mathrm{C}}$ should be compared with
the electron temperature $T_{e}$ which is the measure of the average kinetic energy.
Therefore, it is conventional to describe the electron-electron coupling by
the plasma parameter $\Gamma ^{\left( \mathrm{pl}\right) }=e^{2}\sqrt{\pi
n_{s}}/T_{e}$. For example, the Wigner solid transition occurs~\cite{GriAda-1979}
at $\Gamma ^{\left( \mathrm{pl}\right) }\simeq 131$. The MICO on liquid
helium are usually studied under conditions $\Gamma ^{\left( \mathrm{pl}%
\right) }>10$ when $U_{\mathrm{C}}$ is much larger than the average
kinetic energy. At first glance, it seems that the internal interaction of
such a strength should ruin the quantum picture based on single-electron
Landau levels. Nevertheless, the theoretical treatment using Landau levels
works pretty well because the internal electric field $E_{\mathrm{f}}$ of
fluctuational origin acting on an electron can be considered as a
quasi-uniform electric field~\cite{DykKha-1979}. Such a field can be
eliminated by a proper choice of the reference frame moving along with the
electron orbit center~\cite{MonTesWyd-2002,MonKon-book}. Thus, the 2D Coulomb
liquid can be considered as an ensemble of electrons whose orbit centers are
moving fast in the crossed fields $\mathbf{E}_{\mathrm{f}}$ and $\mathbf{B}$%
. This leads to the Coulomb broadening of the DSF described by Eqs.~(\ref{e15})
and (\ref{e16}). It should be noted, that the quasi-uniform electric field
$\mathbf{E}_{\mathrm{f}}$ does not introduce an additional broadening of
Landau levels because they are defined in the moving
frame, where $\mathbf{E}_{\mathrm{f}}^{\prime }=0$.

\begin{figure}
 \center{ \includegraphics[width=10cm]{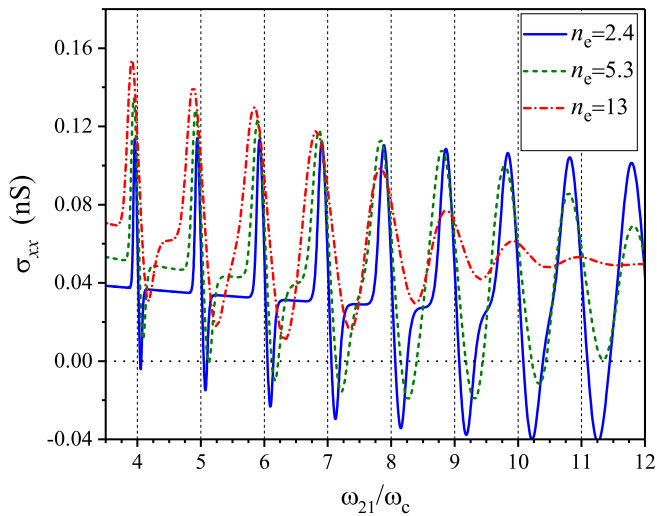}}
\caption{(Color online) $\sigma _{xx}$ vs. $\omega _{2,1}/ \omega _{c} (B)$ calculated for
$T_{e}=T=0.2\,\mathrm{K}$, and for three electron densities $n_{e}$ indicated
in the figure legend in units $10^{6}\,\mathrm{cm}^{-2}$~\cite{KonMonKon-2013}
 }
\label{f8}
\end{figure}

In the theory describing MICO of highly correlated electrons on liquid
helium~\cite{Mon-2012}, the internal field of fluctuational origin $E_{\mathrm{f}}$
is much larger than the driving field $E_{\Vert }$. Therefore, at first, the probability
of scattering is averaged over the fluctuational electric field
$\mathbf{E}_{\mathrm{f}}$ entering the energy conservation
delta-function which leads to the DSF of the 2D Coulomb
liquid~\cite{MonTesWyd-2002,MonKon-book}, and then the probability is expanded
in $eE_{\Vert }\left( X^{\prime }-X\right)$ to obtain the linear conductivity $\sigma _{xx}$.
The main influence of internal interactions on the shape of MICO can be
described by means of the Coulomb broadening parameter $\Gamma _{\mathrm{C}}$
entering $\gamma _{l,n;l^{\prime },n^{\prime }}$ and $\phi _{l,n}$ given in
Eq.~(\ref{e16}). This parameter increases with electron density and electron
temperature as $\Gamma _{\mathrm{C}}\propto n_{e}^{3/4}T_{e}^{1/2}$.
Additionally, the many-electron effect becomes more pronounced at lower
magnetic fields because $\Gamma _{\mathrm{C}}\propto 1/B^{1/2}$. Assuming $%
T_{e}=T=0.2\,\mathrm{K}$, the typical dependence of inter-subband MICO on electron
density predicted by the theory is illustrated in Fig.~\ref{f8}. The blue
solid line calculated for the lowest electron density has rather sharp
variations of $\sigma _{xx}$ near the level matching points with $m\leq 7$.
The broadening of these oscillations steady increases with $m$ (the flat
regions of $\sigma _{xx}$ are shrinking) and at $m\geq 9$ the shape of MICO
is affected by overlapping of sign-changing terms which belong to different
level matching points. The regions with negative conductivity ($\sigma
_{xx}<0$) increase with $m$ for chosen numbers $m<12$.

At higher electron densities presented in Fig.~\ref{f8}, the broadening of
MICO strongly increases because of the Coulomb effect. The regions with $%
\sigma _{xx}<0$ start decreasing with $m\geq 10$ for the olive line ($%
n_{e}=5.3\cdot 10^{6}\,\mathrm{cm}^{-2}$), because $\Gamma _{\mathrm{C}}$ is larger at
lower $B$; they completely disappear for the red line ($%
n_{e}=13\cdot 10^{6}\,\mathrm{cm}^{-2}$). There is also a remarkable prediction of the theory related to
the positions of minima and maxima of the inter-subband MICO. Consider the
shifts of conductivity minima $\delta _{+}=\omega _{2,1}/\omega _{c}\left(
B_{+}\right) -m$ and maxima $\delta _{-}=\omega _{2,1}/\omega _{c}\left(
B_{-}\right) -m$ with regard the level matching point. Here the sign $\pm $
in the subscript means that $\delta _{+}>0$ while $\delta _{-}<0$. Fig.~\ref%
{f8} indicates that the shift of minima $\delta _{+}$ increases
monotonically with $m$, while the shift of maxima has a non-monotonic
dependence clearly seen for olive dashed and red dashed-dotted curves:
after an initial increase, $\left\vert \delta _{-}\right\vert $
attains a maximum value and then decreases strongly for larger $m$. This
is the way the Coulombic corrections to $\gamma _{l,n;l^{\prime
},n^{\prime }}$, and $\phi _{l,n}$ display themselves in the inter-subband MICO.

\begin{figure}
 \center{ \includegraphics[width=10cm]{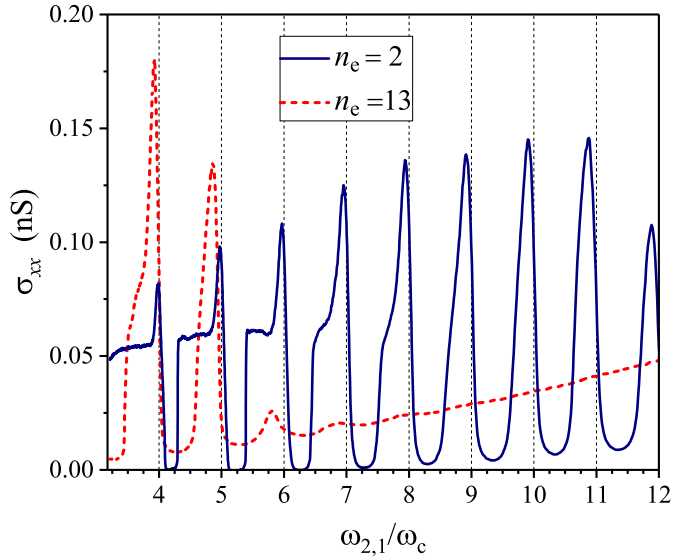}}
\caption{(Color online) $\sigma _{xx}$ vs. $\omega _{2,1}/ \omega _{c} (B)$ obtained at
$T=0.2\,\mathrm{K}$ ($^{3}\mathrm{He}$), input MW power $W=0\,\mathrm{dBm}$, and for two electron densities
$n_{e}$ indicated in the figure legend in units $10^{6}\,\mathrm{cm}^{-2}$~\cite{KonMonKon-2013}
 }
\label{f9}
\end{figure}

The evolution of the shape of $\sigma _{xx}$ oscillations with an increase
in electron density found in the experiment~\cite{KonMonKon-2013} is shown
in Fig.~\ref{f9}. The oscillations are the most pronounced for
the lowest electron density. A noticeable influence of the
Coulomb interaction on $\sigma _{xx}$ oscillations, which magnifies with $m$
, is a strong suppression of the amplitude and an increase in the broadening
of conductivity extrema in accordance with the inter-subband mechanism of
MICO. It should be noted that theoretical lines shown in Fig.~\ref{f8} were
calculated under the assumption $T_{e}=T$. In an experiment with MW
excitation of SE, the elastic decay of electrons to the ground subband is
accompanied by electron heating~\cite{KonMonKon-2007}. Under a magnetic
field, electron decay is possible only near the level matching points.
Therefore, $T_{e}$ oscillates~\cite{Mon-2011} with $\omega _{2,1}/\omega _{c}$
attaining sharp maxima when $\omega _{2,1}/\omega _{c}\simeq m$. These
oscillations of $T_{e}$ make the amplitude of conductivity minima larger
than the amplitude of the respective maxima~\cite{Mon-2011} because the
effective collision frequency caused by intra-subband scattering $\nu _{%
\mathrm{intra}}\propto 1/T_{e}$. This also agrees with experimental
observations~\cite{KonMonKon-2013} shown in Fig.~\ref{f9}. The electron
temperature was estimated to be about $1\,\mathrm{K}$.

\begin{figure}
 \center{ \includegraphics[width=9.5cm]{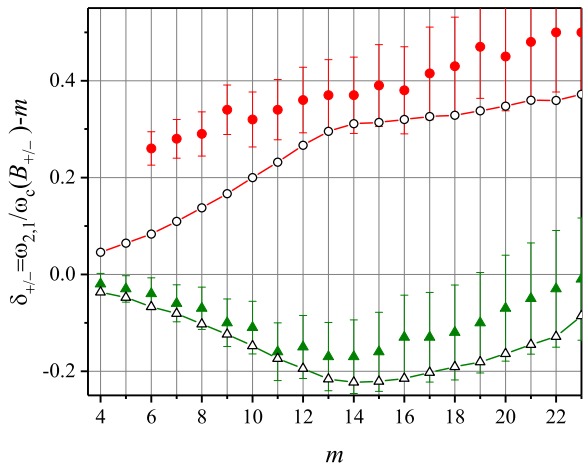}}
\caption{(Color online) Shifts of conductivity minima $\delta _{+}>0$ (circles) and maxima $\delta _{-}<0$
(triangles) vs. the level-matching number $m$ for $n_{e}=2\cdot 10^{6}\,\mathrm{cm}^{-2}$: experimental data (filled symbols),
and the many-electron theory (open symbols with line)~\cite{KonMonKon-2013}
 }
\label{f10}
\end{figure}

As expected, internal forces cause also nontrivial changes in the location
of conductivity extrema. It is instructive to consider positions of
conductivity extrema observed in the experiment versus $m$ , as shown in
Fig.~\ref{f10}. The uncertainty in the positions of oscillation extrema is mostly
determined by the uncertainty in values of $B$. The latter was determined by
\textit{in situ} cyclotron resonance measurements~\cite{KonKon-2010}.
Note that the uncertainty in $\omega _{2,1}/\omega _{c}\propto
B^{-1}$ increases with decreasing $B$. For $m<6$, positions of minima cannot
be determined accurately because of formation of ZRS. The shift of minima $%
\delta _{+}$ (red circles) increases monotonically with $m$ , and at high $m$
the $\delta _{+}\rightarrow 1/2$ which is substantially larger than the same
quantity ($\delta _{+}\simeq 1/4$) reported for intra-subband MIRO in
semiconductor devices~\cite{ManSmeKli-2002}. In contrast, the shift of maxima
$\delta _{-}$ (olive triangles) is non-monotonic. After an
initial increase, $\left\vert \delta _{-}\right\vert $ reaches the maximum
value of about $0.2<1/4$ at $m=13$, and decreases strongly for larger $m$.
Calculations shown in Fig.~\ref{f10} were performed for the MW field
resulting in conductivity oscillations of approximately the same amplitude
as in the experiment. Quantitative differences between data and the theory
(open symbols) can be attributed to electron heating. Nevertheless, the
delicate theoretical findings, which concern the difference in behavior of
positions of conductivity extrema as functions of electron density and of
the level-matching number $m$, are clearly observed in the
experiment. Thus, the behavior of positions of conductivity extrema observed
is drastically different from the behavior of conductivity extrema in
GaAs/AlGaAs heterostructures, in accordance with predictions of the
inter-subband theory of MICO.

\begin{figure}
 \center{ \includegraphics[width=9cm]{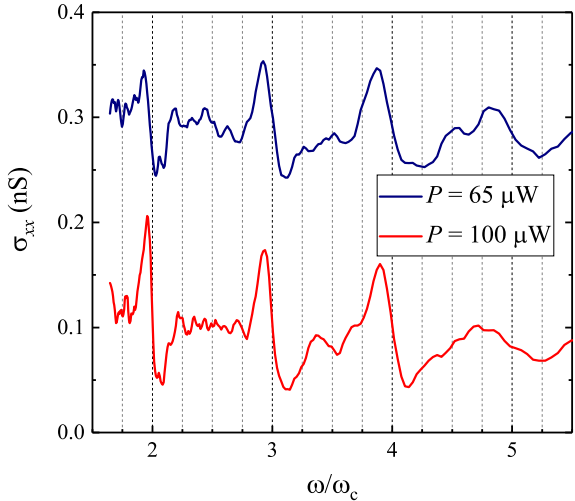}}
\caption{(Color online) $\sigma _{xx}$ vs. $\omega / \omega _{c} (B)$ obtained at
$T=0.56\,\mathrm{K}$, $\omega /2\pi =88.52\,\mathrm{GHz}$, $n_{e}=1.7\cdot 10^{7}\,\mathrm{cm}^{-2}$, and different
values of the incident MW power $P$. For clarity, the curve
for $65\,\mu \mathrm{W}$ (dark blue) is
upshifted by $0.2\,\mathrm{nS}$~\cite{YamMonKon-2015}
 }
\label{f11}
\end{figure}

Regarding the MICO of SE caused by intra-subband photon-assisted scattering,
the comparison between the experiment and theory shown in Fig.~\ref{f4}
indicates that Coulomb broadening $\Gamma _{\mathrm{C}}$ can correctly
describe the width of oscillatory features of $\sigma _{xx}$. The
experimental data~\cite{YamMonKon-2015} shown in Fig.~\ref{f11} clear indicate
that the shifts of conductivity minima $\delta _{+}=\omega /\omega
_{c}\left( B_{+}\right) -m$ and maxima $\left\vert \delta _{-}\right\vert
=\left\vert \omega /\omega _{c}\left( B_{-}\right) -m\right\vert $ steady
increase with $m$, at least up to $m=5$ where they reach the number $1/4$.
Thus, for high magnetic fields, oscillatory variations are strongly confined
near $\omega /\omega _{c}=2,3,$ and $4$, and positions of minima are not
fixed to "magic" numbers $m+1/4$ which is
in contrast with data obtained for GaAs/AlGaAs in Ref.~\cite{ManSmeKli-2002}. This
difference can be attributed to a substantially smaller collision broadening
of Landau levels of SE on liquid helium as compared to that of semiconductor
devices. It should be noted also that deviations from a "$1/4$-cycle shift" are found
in the ZRS regime ($m\leq 4$) even for semiconductor electrons~\cite{Zud-2004}.

\section{ Resonant Photovoltaic Effect}
\label{sec:4}

In the MICO experiments~\cite{KonKon-2009,KonKon-2010} based on resonant
inter-subband excitation, the
magnetoconductivity data were obtained by measuring the average response of
the electron system to a driving in-plane electric field. A remarkable
result was found~\cite{KonCheKon-2012} by detecting the photoresponse of
surface electrons in the absence of the driving electric filed under the
conditions where dissipative conductivity is vanishing $\sigma _{xx}$ $%
\rightarrow 0$. The ultra-strong photovoltaic effect observed in this
experiment is characterized by a nonequilibrium spatial distribution of
electrons in the confining electrostatic potential. Moreover, the
electrostatic energy acquired by an electron exceeds other relevant energies
by several orders of magnitude.

\begin{figure}
 \center{ \includegraphics[width=9cm]{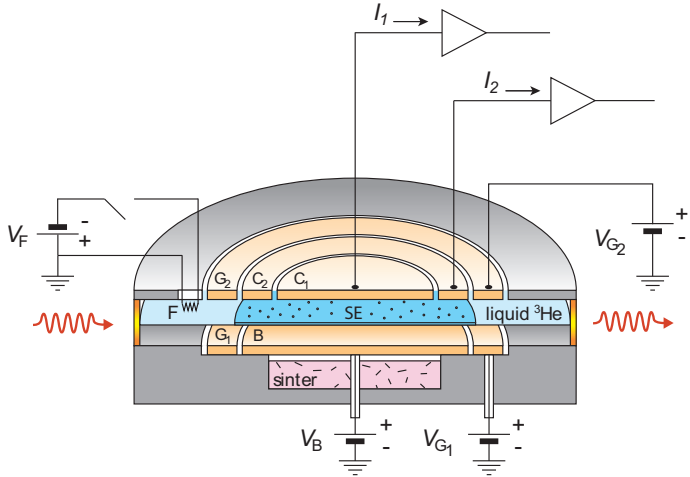}}
\caption{(Color online) Schematic diagram of the experimental method.
Detailed description is provided in the text~\cite{KonCheKon-2012}
 }
\label{f12}
\end{figure}

\begin{figure}
 \center{ \includegraphics[width=8cm]{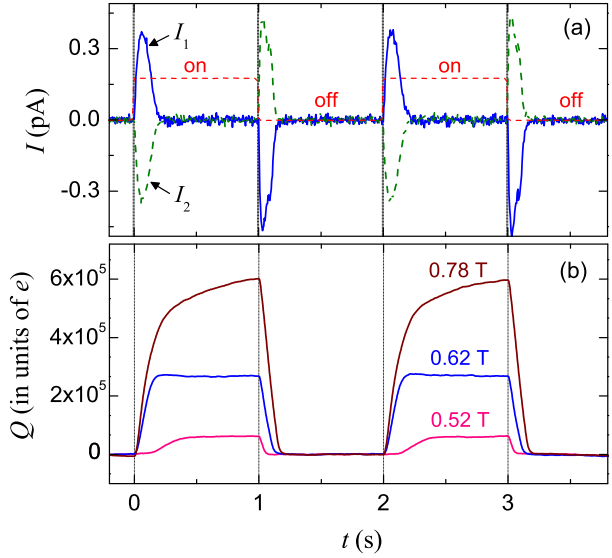}}
\caption{(Color online) (a) Transient signals of photocurrents $I_1$ (solid line,
blue) and $I_2$ (dashed line, green) induced in electrodes $C_1$ and $C_2$,
respectively, by the flow of the surface charge at $T=0.2\,\mathrm{K}$ and $B=0.62\,\mathrm{T}$.
Short dashed line (red) is a square waveform, which switches the MW
source on (off) at a high (low) signal level. (b) Cumulative charge $Q$
obtained by integrating the current $I_1$ at three values of $B$ corresponding to
$m=4$ ($0.78\,\mathrm{T}$), $m=5$ ($0.62\,\mathrm{T}$), and $m=6$ ($0.52\,\mathrm{T}$)
conductance minima~\cite{KonCheKon-2012}
 }
\label{f13}
\end{figure}

Redistribution of SE was detected by measuring photocurrents $I_{1}$ and $%
I_{2}$ induced in the inner (C$_{1}$) and outer (C$_{2}$) electrodes of a
Corbino disk placed just above the electron pool, as shown in Fig.~\ref{f12}.
In the absence of MW radiation, potentials applied to the guard electrodes G$%
_{1}$ and G$_{2}$, and to the bottom disk ($V_{\mathrm{B}}$) form a nearly
uniform electron density along the liquid helium surface with a sharp edge.
Electrons were tuned for the inter-subband resonance $\omega =\omega _{2,1}$
with the applied MW by adjusting $V_{\mathrm{B}}$. In order to study
the transient photoresponse of SE, the incident MW power was pulse
modulated using a low-frequency (0.5--6 Hz) square waveform. The results of
measurements are shown in Fig.~\ref{f13} for currents $I_{1}$ and $I_{2}$
and the cumulative charge $Q$. The value of the magnetic field used for obtaining the current
data shown in Fig.~\ref{f13}(a) corresponds
to the $m=5$ conductance minimum. The modulation of the MW power by
the square waveform is shown in Fig.~\ref{f13}(a) by a red dashed line. Sharp
changes in $I_{1}$ and $I_{2}$ observed indicate that, upon switching the
power on, the electrons are pulled by radiation towards the edge of the
electron pool, causing the depletion of the charge in the central region of
the pool. Correspondingly, the positive (negative) current is induced in
electrode C$_{1}$ (C$_{2}$) by the flow of the image charge. The surface
charge flows until a new spatial distribution of electrons in the unchanged
confining electrostatic potential is established, after which the currents $%
I_{1}$ and $I_{2}$ becomes zero. Because of the displacement of SE with
respect to the neutralizing background, a non-zero electric field is
developed in the charged layer. Upon switching the power off, the displaced
surface charge flows back to restore the equilibrium distribution of
electrons. Correspondingly, a negative (positive) current of the image
charge is induced in C$_{1}$ (C$_{2}$).

The cumulative charge $Q$ flowing from, for example, the electrode C$_{2}$
is obtained by integrating the measured current $I_{1}$. The $Q$ is shown in
Fig.~\ref{f13}(b) in units of the elementary charge ($e>0$) for three values
of $B$ corresponding to the conductance minima $m=4$, $5$, and $6$. The
estimation given in Ref.~\cite{KonCheKon-2012} indicates that a very large
fraction (more than 50\%) of the surface charge can be displaced upon
irradiation. The numerical calculations show that the displacement of 50\%
of electrons leads to the potential difference between the central and
peripheral parts $V_{e}\approx 0.3\,\mathrm{V}$. This result corresponds to
the increase in electrical potential energy of a single electron exceeding
other relevant energy scales such as, for example, the inter-subband energy
difference or $T_{e}$, by several orders of magnitude.

\begin{figure}
 \center{ \includegraphics[width=9cm]{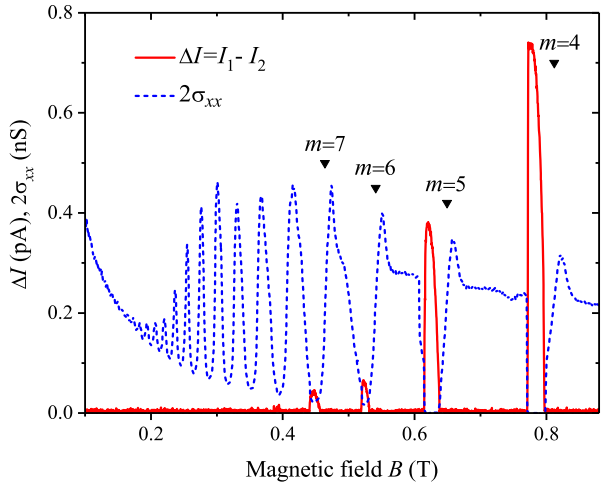}}
\caption{(Color online) Current signal $\Delta I$ (red solid) obtained under 100\% modulation of the
MW for $n_{e}=1.4\cdot 10^{6}\,\mathrm{cm}^{-2}$ and $T=0.2\,\mathrm{K}$ is compared with
MICO (blue dashed) obtained under the same conditions. Black triangles indicate the values
of $B$ such that $\omega _{2,1}/\omega _c=m$~\cite{KonCheKon-2012}
 }
\label{f14}
\end{figure}

A comparison between $\sigma _{xx}$ measured under MW irradiation and $%
\Delta I=I_{1}-I_{2}$ recorded under 100\% modulation of the incident
MW power is shown in Fig.~\ref{f14}. Both sets of data were obtained%
~\cite{KonCheKon-2012} under the same experimental conditions and at the
same level of MW power. The Fig.~\ref{f14} proves a clear relationship
between the conductance minima and the transient current: a nonzero signal $%
\Delta I$ and a displacement of the surface change are observed only in
the intervals of $B$ near the conductance minima corresponding to $m=4$, $5$%
, $6$, and $7$. No signal is observed when electrons are tuned away from the
inter-subband resonance by changing the electrical bias $V_{\mathrm{B}}$.
Simultaneously, the detuning results in the complete disappearance of the
MICO and ZRS.

\begin{figure}
 \center{ \includegraphics[width=9cm]{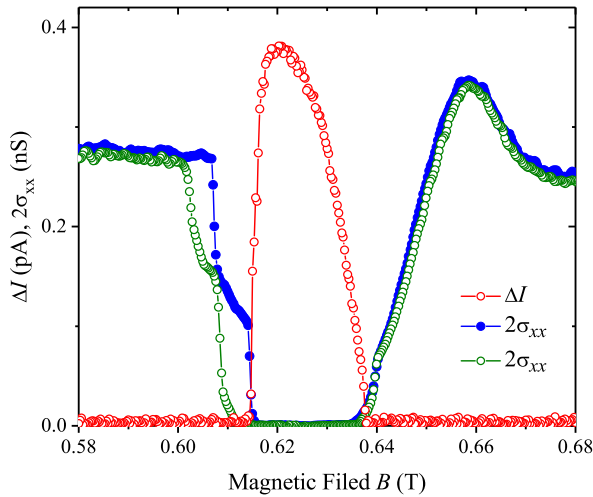}}
\caption{(Color online)  $\Delta I$ (red) in the range of $B$ close to $m=5$. For
comparison, $2\sigma _{xx}$ is shown for upward (solid circles) and
downward (open circles) sweeps of $B$. Experimental conditions are the same as in Fig.~\ref{f14}~\cite{KonCheKon-2012}
 }
\label{f15}
\end{figure}

A detailed relationship between $\sigma _{xx}$ and $\Delta I$ in a narrow
range of $B$ near the $m=5$ conductance minimum\ is illustrated in
Fig.~\ref{f15}. The signal $\Delta I$ emerges sharply upon slowly increasing $B$ when
$\sigma _{xx}$ drops to zero. Thus, the abrupt change of $\Delta I$ is an
indication of instability of the electron system leading to formation of
ZRS. Upon the downward sweep of $B$, $\sigma _{xx}$ (open circles) exhibits
hysteresis. Such hysteresis is a feature of a metastable state coexisting
with the global stable state of the electron system. We expect that such a
hysteresis can be caused by decay heating of SE, which requires an
additional theoretical investigation in the ripplon dominated scattering regime.

It is reasonable to attribute the resonant photovoltaic phenomena observed%
~\cite{KonCheKon-2012} to the effect of absolute negative dissipative
conductivity ($\sigma _{xx}<0$) which appears in the inter-subband mechanism
of MICO~\cite{Mon-2011,Mon-2012}. For example, in semiconductor systems, ZRS
are explained~\cite{AndAleMil-2003} as a consequence of the negative linear
conductivity condition $\sigma _{xx}<0$ which appears for high radiation
power. Due to instability of the system under this condition, it enters a
nonlinear regime and develops a steady current state $j_{0}$ with $\sigma
_{xx}(j_{0})=0$. This predicts the existence of current domains, where
electrons move in opposite directions~\cite{AndAleMil-2003}. Such a model of
the ZRS is well applicable for semiconductor electrons. By the way, the
current flow anomalies (similar to the photovoltaic effect discussed here)
reported for the irradiated 2D electron gas in semiconductor
structures~\cite{WilPheWes-2004} were ascribed to the theoretical pictures of
instabilities due to local negative resistivities.

In the 2D Coulomb liquid on the surface of liquid helium, formation of
current domains is unlikely because of strong electron correlations ($\Gamma
^{\left( \mathrm{pl}\right) }\gg 1$). The system of SE usually have no source
and drain electrodes. Therefore, a steady current can be formed only by
electrons circling the center of the electron pool. Since it is impossible
to create a strong current density in the center, electrons will move to the
edge of the electron liquid depleting the center. As a result, a nonuniform
electron density distribution along the surface will be formed to provide
radial electric field and a circling current, strong enough to make $\sigma
_{xx}=0$. At a fixed magnetic field, a change in electron density
additionally helps the system to leave the unstable regime, due to the
Coulombic effect (see also related discussions in Chapter~\ref{sec:7}). It
should be noted here that small transient photocurrents are also observed
near the average conductivity minima corresponding to $m=6$ and $7$ which do
not reach zero. This can be explained by the assumption that under these
conditions only a part of electrons has $\sigma _{xx}<0$ and moves uphill in
the confining potential. Remarkably, for the $m=6$ minimum with $\sigma
_{xx}>0$, a time delay of up to $0.1\,\mathrm{s}$ between the application of
microwaves and the onset of the charge motion is observed, as indicated in
Fig.~\ref{f13}(b), which is in contrast with the results obtained for regions
with $\sigma _{xx}=0$. The inter-subband mechanism of MICO and absolute
negative conductivity do not require the in-plane component of the MW field.
It should be noted that the in-plane component of $\mathbf{E}_{\mathrm{mw}}$ can also
be a reason for an additional photocurrent~\cite{EntMag-2013}, though its
relation to the conductance minima is less evident.

\section{Self-generated audio-frequency oscillations}
\label{sec:5}

In the regime of vanishing diagonal (dissipative) conductivity $\sigma
_{xx}\rightarrow 0$, in addition to the static pattern (a strong depletion
of charge at the center of the electron layer), the redistributed charge
exhibits spontaneously generated oscillations in the audio-frequency range%
~\cite{KonWatKon-2013}. Oscillations reported were observed as an electrical
current $I$ induced in a circular metal electrode $\mathrm{C}_1$ ($7\,\mathrm{mm}$ radius),
which was a part of the Corbino disk located $1.3\,\mathrm{mm}$ above the surface
and used also for conductivity measurements (see Fig.~\ref{f12}).

\begin{figure}
 \center{ \includegraphics[width=8cm]{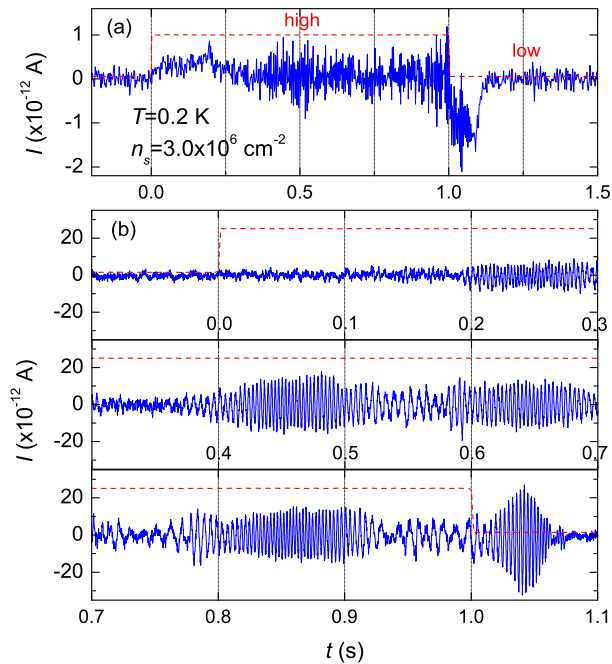}}
\caption{(Color online) Current at a metal electrode capacitively
coupled to 2D electrons on the surface of liquid $^{3}\mathrm{He}$. The data were obtained at
$T=0.2\,\mathrm{K}$ and $B=0.62\,\mathrm{T}$ using microwaves of frequency $\omega/2\pi=90.9\,\mathrm{GHz}$. The
gain of current preamplifier sets its bandwidth, which was $100$ and $2000\,\mathrm{Hz}$ for
the data shown in panel (a) and (b), respectively. The high (low) level of the
square waveform (dashed line) corresponds to microwave power on (off)~\cite{KonWatKon-2013}
 }
\label{f16}
\end{figure}

The transient current triggered by switching on and off radiation pulses and
shown above in Fig.~\ref{f13}(a) is actually an average over many traces. In
a single trace, the current $I$ shown in Fig.~\ref{f16}(a) has also
oscillations which appear spontaneously. MW radiation was applied
during $0\leq t\leq 1.0\,\mathrm{s}$, as indicated by the square red dashed
line. The broad peak at $0< t < 0.25\,\mathrm{s}$ was due to
the depletion of charge at the center of the pool as electrons were moving
toward the edge. A peak of opposite sign at $1.0< t < 1.1%
\,\mathrm{s}$ appeared as microwaves were switched off and the system restored
the equilibrium charge distribution. According to Fig.~\ref{f16}(a) there
exists an additional oscillating signal at $0.25\leq t\leq 1.1%
\,\mathrm{s}$, that persists as long as the system retains nonequilibrium
distribution.

The trace of Fig.~\ref{f16}(a) was obtained similar to the
Ref.~\cite{KonCheKon-2012} using a current preamplifier with a bandwidth of $100%
\,\mathrm{Hz}$, which attenuated significantly the high-frequency content of
the current oscillations. This filtering was avoided by increasing the
bandwidth to $2\,\mathrm{kHz}$. The respective time sequence of recorded
oscillations is shown in Fig.~\ref{f16}(b). In this case, the amplitude of
oscillating current was much larger than the current induced due to charge
redistribution. The latter was too small to be seen in Fig.~\ref{f16}(b)
because of the reduced sensitivity of the current preamplifier. The charge
displaced during the half-period of the oscillations at their maximum
amplitude was estimated to be about 5\% of the total surface charge.

\begin{figure}
\center{\includegraphics[width=10cm]{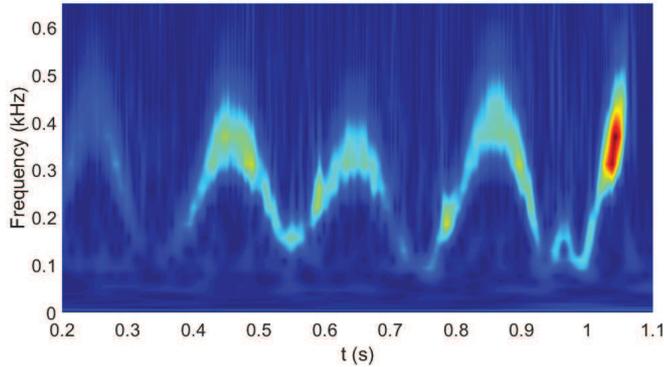}}
\caption{(Color online) Wavelet transform of oscillations shown in
Fig.~\ref{f16}. Color represents intensity of the spectral component ($Y$-axis) of
oscillations at a given time $t$~\cite{KonWatKon-2013}
 }
\label{f17}
\end{figure}

An important feature of the persisting oscillations shown in Fig.~\ref{f16}%
(b) is that they are not monochromatic. It is convenient to look at the
wavelet transform, which provides information about the instantaneous
frequency content of the recorded signal. The wavelet transform for the
oscillations in Fig.~\ref{f16}(b) is shown in Fig.~\ref{f17}. The frequency varies
periodically in time in the range of 100--500\,Hz. These frequencies are in
the audible range, which makes it possible to hear them by ear. In most
experiments done with electrons on liquid $^{3}\mathrm{He}$, as well as
superfluid $^{4}\mathrm{He}$, similar periodic variation of the frequency of
oscillations with a period of about $0.2\,\mathrm{s}$ (independent of the
substrate) were found. At the same time, the frequency of self-generated
oscillations is about twice as large for electrons on superfluid $^{4}%
\mathrm{He}$ comparing with electrons on liquid $^{3}\mathrm{He}$.

It is well known~\cite{CraWil-1972} that a 2D electron pool and the perpendicular
electric field $E_{\bot }$ applied put pressure on the free surface of
liquid helium causing a steady surface deformation proportional to the electron
density and $E_{\bot }$. The depth of the surface depression $h$ is substantial and can even be about $%
0.1\,\mathrm{mm}$ (at high $n_{e}$ and $E_{\bot }$). Obviously, the
redistribution of the electron charge caused by the photovoltaic effect should induce a change in the surface
depression reducing it in the center and increasing it in edge regions.
For experimental conditions of Ref.~\cite{KonWatKon-2013}, variations of the depth
$\delta h$ induced by the redistribution of electrons can be about $10^{-6} \mathrm{cm}$.
Therefore, the electron pool of displaced SE acquires a huge inertia.
These changes of $h$ in the central and edge regions cannot
detune the electron system from the inter-subband resonance because respective
corrections to $\Delta _{2,1}$ are about four orders of magnitude smaller than the
typical line-width of the resonance caused by inhomogeneity of the electric field:
$2\gamma _{\mathrm{mw}}\simeq 0.2\,\mathrm{GHz}$. Nevertheless, one can expect that
inhomogeneity is larger in the edge region than in the central part of the electron pool.
In this case, even a small increase in $\gamma _{\mathrm{mw}}$ can detune slightly
displaced SE from the inter-subband resonance in the edge region and eliminate the negative conductivity.
The large deformation relief has a huge inertia leading to a delay before these electrons
start moving back tuning themselves to the resonance again. This effect can cause radial
oscillations of electron density coupled with gravity waves. The lowest
frequency of radially symmetrical mode of the gravity wave has angular
frequency~\cite{KonWatKon-2013} $\omega \approx 3.83\sqrt{gH}/R$ (here $R$ is
the radius of the electron pool, $H$ is the height of the liquid surface,
and $g$ is the acceleration due to gravity) corresponding to about $3.6%
\,\mathrm{Hz}$, which is close to the observed periodic variation of the
self-generated frequency. Possible reasons for self-generated oscillations will be
discussed in the Section~\ref{sec:7}.

\section{Incompressible states}
\label{sec:6}

A striking example of irradiation-induced self-organization was
observed~\cite{CheWatKon-2015} in a coupled system of two electron gases of different
densities. The 2D electron gases were formed on the surface of liquid $^{4}%
\mathrm{He}$: near the central Corbino electrode (with density $n_{\mathrm{e}%
}$), and near the guard-ring electrode (with density $n_{\mathrm{g}}$). The
Corbino disk was similar to that shown in Fig.~\ref{f12}. Density of electrons coupled to the central
electrode $n_{\mathrm{e}}\left( r\right) $ was approximately uniform.
Regarding $n_{\mathrm{g}}\left( r\right) $, it was strongly nonuniform
having a sharp maxima near the middle of the guard strip. Therefore, the
notation $n_{\mathrm{g}}$ was attributed to the mean electron density in the
guard region. At a fixed total number of electrons $N_{e}$, the ratio $n_{\mathrm{e}%
}/n_{\mathrm{g}}$ was varied by changing the potential $V_{\mathrm{g}}$
applied to the guard electrode. In the presence of resonant MW radiation,
under the magnetic field fixed to the ZRS condition $\omega _{2,1}/\omega
_{c}\left( B\right) =6.25$, the inner 2D electron gas enters an
incompressible state with an electron density $n_{\mathrm{e}}=n_{\mathrm{c}%
}\simeq 3.4\cdot 10^{6}\,\mathrm{cm}^{-2}$ independent of $N_{e}$ and of the
potential applied to the guard electrode for a wide range of parameters.

The compressibility of the 2D electron gas was defined as~\cite{CheWatKon-2015}%
\begin{equation}
\chi =-\frac{dn_{\mathrm{e}}}{dV_{\mathrm{g}}}.  \label{e37}
\end{equation}%
In the absence of MW irradiation, over a large range of $V_{\mathrm{g}}$ and
$N_{e}$, the compressibility was well approximated by the constant value $%
\chi _{0}\simeq 2.9\cdot 10^{6}\,\mathrm{cm}^{-2}\mathrm{V}^{-1}$ in a good
agreement with estimations obtained using the parameters of the experimental
cell.

\begin{figure}
 \center{ \includegraphics[width=9cm]{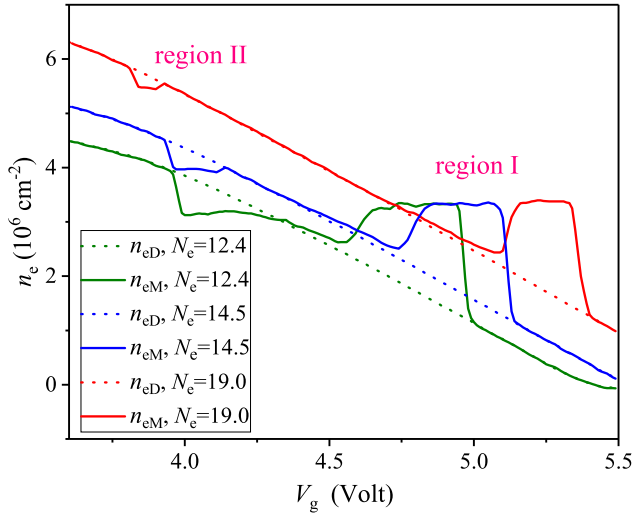}}
\caption{(Color online) Density of electrons coupled to the central
electrode $n_{\mathrm{e}}$ vs. the potential of the guard electrode
$V_{\mathrm{g}}$ for $N_{e}=12.4, 14.5, 19\,[\times 10^{6}]$:
in the dark $n_{\mathrm{eD}}$ (dotted) and under MW irradiation
$n_{\mathrm{eM}}$ (solid)~\cite{CheWatKon-2015}
 }
\label{f18}
\end{figure}

Outside the ZRS regions the compressibility $\chi $ was
not affected by MW irradiation because it is independent of $\sigma _{xx}$
for a stable regime ($\sigma _{xx}>0$). In a ZRS region, a remarkable change
of $\chi $ was reported. The experimental dependence of $n_{\mathrm{e}%
}\left( V_{\mathrm{g}}\right) $ is shown in Fig.~\ref{f18} for different $N_{e}$. In
the dark case (dotted lines marked $n_{\mathrm{eD}}$ in the Legend),
electron compressibility defined by Eq.~(\ref{e37}) is nearly constant.
Under MW radiation in two distinct regions (I and II), the
dependence $n_{\mathrm{e}}\left( V_{\mathrm{g}}\right) $ changes
drastically, as shown by the solid lines marked $n_{\mathrm{eM}}$. For example,
consider the blue solid line ($N_{e}=14.5\cdot 10^{6}$).  Below $5.13%
\,\mathrm{V}$ there is a sharp (nearly vertical) increase of $n_{\mathrm{eM}}$
up to a value $n_{\mathrm{c}}\simeq 3.4\cdot 10^{6}\,\mathrm{cm}^{-2}$. Then,
there is a plateau with $\chi =0$. Remarkably, in the range $4.75\,\mathrm{V}%
<V_{\mathrm{g}}<4.9\,\mathrm{V}$ the system exhibits a negative (!)
compressibility: $\chi <0$. At lower $V_{\mathrm{g}}$, the $n_{\mathrm{eM}%
}\left( V_{\mathrm{g}}\right) $ returns to the dependence observed for the
dark case $n_{\mathrm{eD}}\left( V_{\mathrm{g}}\right) $. The value $n_{%
\mathrm{c}}$ is not a universal constant because it depends on the
conductivity minimum chosen: $n_{\mathrm{c}}$ decreases with the level
matching number $m$. It was noted that for the magnetic field corresponding
to $\omega _{2,1}/\omega _{c}=10+1/4$ the position of $n_{\mathrm{c}}$ is
displaced towards significantly lower densities $n_{\mathrm{c}}\simeq
1.3\cdot 10^{6}\,\mathrm{cm}^{-2}$.

The negative compressibility observed means that the electron pool squeezes
when a potential gradient stretches it, and the pool expands when a potential
gradient compresses it. This agrees with our understanding of negative conductivity
effects: in an external electrostatic potential electrons flow uphill. It
should be noted that already from the existence of regions with $\sigma
_{xx}<0$ ($\chi <0$) and $\sigma _{xx}>0$ ($\chi >0$) it follows that there
should be at least one incompressible state ($\chi =0$) in between them. The
most puzzling thing is that the incompressible state is observed in a quite
broad region of $V_{\mathrm{g}}$. Upon a decrease of the total number of SE ($%
N_{e}$), the plateau of Fig.~\ref{f18}, as well as the regions with $\chi
=\chi _{0}$, moves to a lower $V_{\mathrm{g}}$, while its vertical position
remains unchanged: $n_{\mathrm{c}}\simeq 3.4\cdot 10^{6}\,\mathrm{cm}^{-2}$.

At $V_{\mathrm{g}}\approx 4\,\mathrm{V}$ (region II), there is another state with $\chi
\rightarrow 0$, however, the plateau density value depends on $N_{e}$ in
contrast to the region I ($V_{\mathrm{g}}\approx 5\,\mathrm{V}$). It is reasonable
to ascribe this plateau to an incompressible state of electrons situated
above the guard ring, because the plot $n_{\mathrm{gM}}\left( V_{\mathrm{g}%
}\right) $ (not shown here) indicates that the vertical position of the corresponding plateau
($V_{\mathrm{g}}\approx 4\,\mathrm{V}$) is practically unchanged giving
another critical density $n_{\mathrm{gM}}=n_{\mathrm{c}}^{\left( 2\right)
}\simeq 1\cdot 10^{6}\,\mathrm{cm}^{-2}$. The relative "weakness" of this
second incompressible state may be caused by the initial nonuniformity of
the electron density near the guard-ring electrode $n_{\mathrm{g}}\left(
r\right) $. Anyway, a mechanism selecting a particular density value was not
explained in the original paper~\cite{CheWatKon-2015}. Possible relationship of
the observed phenomenon with absolute negative conductivity and the Coulombic
effect on MICO will be discussed in the next Section.

\section{Density domains}
\label{sec:7}

A state of an electron system with $\sigma _{xx}<0$ is unstable, and usually
a certain local current density $j_{0}$ is necessary to reach the stable
state $\sigma _{xx}\left( j_{0}\right) =0$ owing to nonlinear
effects~\cite{AndAleMil-2003}. According to this theory, the instability of a 2D electron
system caused by negative conductivity leads to the pattern of the current
distribution in a conducting strip: domain wall separating currents of the
same amplitude $j_{0}$ but opposite directions. The position of the domain
wall determines the net current. The current pattern in the Corbino geometry
is obtained by connecting the edges of the current strip into a ring.

Surface electrons on liquid helium represent a highly correlated system,
where the average Coulomb interaction potential energy per an electron is
much larger than the average kinetic energy. In such an electron liquid,
current domains are unlikely due to the strong mutual friction of currents
in the domain wall (the electron system is close to the Wigner solid state).
Therefore, the internal structure of the ZRS of SE on liquid helium requires
an additional investigation. Considering only nonlinear effects of $\sigma
_{xx}\left( j\right) $, as the origin of stability in a Corbino geometry
(without current domains), we came to the conclusion~\cite{Mon-2012} that the
center of the electron pool should be depleted in accordance with
experimental observation of the ultra-strong photovoltaic effect~\cite{KonCheKon-2012}.
A strong displacement of electrons against the confining
force of Corbino electrodes gives us an important insight into this problem
because the Coulombic effect on magnetoconductivity minima discussed in
Section~\ref{sec:3} can also be the cause making $\sigma _{xx}\geq 0$.

At first, it is instructive to track the evolution of $\sigma _{xx}$ minima with
increasing SE density $n_{e}$ qualitatively using simple analytical equations.
According to Eq.~(\ref{e34}), the integrand of the expression for $\sigma _{xx}$
contains derivatives of the sum of Gaussian functions entering the DSF of the Coulomb liquid.
If we neglect overlapping of the sign-changing terms in the sum over the all $m$,
then positions of minima of the integrand are given by%
\begin{equation}
\frac{\omega _{2,1}}{\omega _{c}}=m+\frac{\gamma _{2,0;1,m}}{\sqrt{2}\omega
_{c}}+x_{q}\frac{\Gamma _{\mathrm{C}}^{2}}{4T_{e}\hbar \omega _{c}},
\label{e38}
\end{equation}%
where $\gamma _{l,n;l^{\prime },n^{\prime }}$ defines the broadening of $%
S_{l,l^{\prime }}\left( q,\Omega \right) $ minima. The last term in the right side
represents the Coulomb correction which depends on the wavevector $\mathbf{q}$. Therefore,
integration over $\mathbf{q}$ affects the position of conductivity minima. Numerical
calculations indicate that for estimation of positions of $\sigma _{xx}$ minima
we can use Eq.~(\ref{e38}) with a simple replacement $x_{q}\sim m$ (the results of accurate
numerical calculations are shown below in Fig.~\ref{f19}). Thus, the shift of a conductivity minimum from an
integer $m$ increases monotonically with electron density due to the Coulomb broadening $%
\Gamma _{\mathrm{C}}$ entering also $\gamma _{2,0;1,m}$. At large $m$ and $%
\Gamma _{\mathrm{C}}$ when the sign-changing terms become strongly
overlapping, the positions of minima are asymptotically given by
\begin{equation}
\frac{\omega _{2,1}}{\omega _{c}}=m+\frac{1}{4}+x_{q}\frac{\Gamma _{\mathrm{C%
}}^{2}}{4T_{e}\hbar \omega _{c}},  \label{e39}
\end{equation}%
where $x_{q}\sim m$. In this limiting case, the deviation of minima from the level matching point
also increases monotonically with electron density.

\begin{figure}
 \center{ \includegraphics[width=9cm]{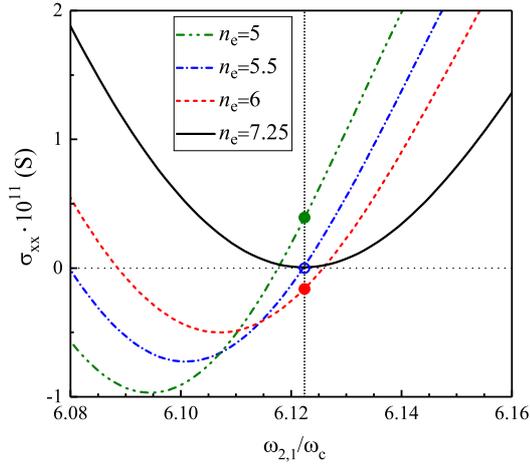}}
\caption{(Color online) Magnetoconductivity in a dc electric field calculated for
$T=0.3\,\mathrm{K}$ (liquid $^{4}\mathrm{He}$) and four values of $n_{e}$. Values of $n_{e}$ are shown in units of $10^{6} \,\mathrm{cm}^{-2}$.
Circles indicate evolution of $\sigma _{xx}$ with varying $n_{e}$ for a fixed $B$ (vertical line)~\cite{Mon-2016}
 }
\label{f19}
\end{figure}

For a typical region of $B$, the evolution of the position of a conductivity minimum
with varying $n_{e}$ is illustrated in Fig.~\ref{f19}.
One can see that an increase in $n_{e}$ affects the chosen
minimum in two ways: the position of the minimum moves to the region of larger $%
\omega _{2,1}/\omega _{c} $ and higher $\sigma _{xx}$. Therefore, at a fixed magnetic field shown by
the vertical dotted line, and at the lowest density $n_{e}=5\cdot 10^{6}\,\mathrm{%
cm}^{-2}$, electron conductivity $\sigma _{xx}>0$, as indicated by the olive
circle representing the cross point of the olive dash-dot-dotted line with
the vertical line. The increase of $n_{e}$ up to $5.5\cdot 10^{6}\,\mathrm{cm}%
^{-2}$ moves the cross point down making the ZRS ($\sigma _{xx}=0$), as
indicated in Fig.~\ref{f19} by the open blue circle. A further increase in $%
n_{e}$ makes absolute negative conductivity ($\sigma _{xx}<0$) shown by the
red circle representing the cross point of the red dashed curve ($%
n_{e}=6\cdot 10^{6}\,\mathrm{cm}^{-2}$) and the vertical line. The vertical
shift of the minimum caused by the Coulombic effect eventually makes $\sigma
_{xx}\geq 0$, as shown by the solid curve calculated for $n_{e}=7.25\cdot
10^{6}\,\mathrm{cm}^{-2}$. Thus, for a fixed magnetic field, the instability ($%
\sigma _{xx}<0$) appears inside a certain region restricted by two critical
densities (higher $n_{\mathrm{H}}$ and lower $n_{\mathrm{L}}$). At $n_{e}=n_{%
\mathrm{L}}$, or at $n_{e}=n_{\mathrm{H}}$\ the system is in the ZRS. The
length of the unstable region $n_{\mathrm{H}}-n_{\mathrm{L}}$ depends
strongly on the position of the vertical line (chosen $B$) with regard to
the ZRS appeared for the solid curve of Fig.~\ref{f19}. It should be noted
that theoretical calculations shown in this figure were calculated assuming $%
T_{e}=T$. Heating of SE induced by decay of the excited electrons increases
the fluctuational electric field and the Coulombic effect; therefore, it
should decrease the estimated values of $n_{\mathrm{H}}$ and $n_{\mathrm{L}}$%
.

A negative $\sigma _{xx}$ means that any density fluctuation $\delta n_{e}$
(positive or negative) diffusively grows, because it induces a potential
gradient for electrons moving uphill. This is the natural reason for the
instability. It is quite obvious that the density of growing regions is
limited by the conditions: $n_{e}+\delta n_{e}=n_{\mathrm{H}}$ and $%
n_{e}+\delta n_{e}=n_{\mathrm{L}}$ (for $\delta n_{e}<0$). Indeed, a change
of $n_{e}+\delta n_{e}$ above $n_{\mathrm{H}}$ (below $n_{\mathrm{L}}$)
makes $\sigma _{xx}>0$ and the excess charge will move back in the field
induced by the fluctuation. Therefore, the electron system with $n_{\mathrm{L%
}}<n_{e}<n_{\mathrm{H}}$\ eventually will be separated into fractions
(domains) with different densities $n_{\mathrm{L}}$ and $n_{\mathrm{H}}$.

\begin{figure}
 \center{ \includegraphics[width=9cm]{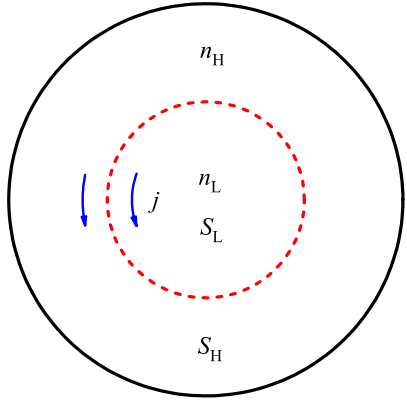}}
\caption{(Color online) The pattern of the density distribution with a domain wall for the
Corbino geometry~\cite{Mon-2016}
 }
\label{f20}
\end{figure}

For a Corbino geometry, the simplest stable pattern of the density
distribution with a domain wall is shown in Fig.~\ref{f20}. It is assumed
that the direction of the charge displacement caused by negative
conductivity should be opposite to the direction of the confining force of
Corbino electrodes acting on SE. Two blue arrows indicate that in different
domains near the domain wall (red dashed circle) local currents flow in the
same direction in contrast with the case of current domains. The position of
the domain wall (the areas $S_{\mathrm{H}}$ and $S_{\mathrm{L}}$)
corresponding to the initial density $n_{e}$ is determined by the simple rule%
~\cite{Mon-2016}:%
\begin{equation}
\frac{S_{\mathrm{H}}}{S_{\mathrm{L}}}=\frac{n_{e}-n_{\mathrm{L}}}{n_{\mathrm{%
H}}-n_{e}}  \label{e40}
\end{equation}%
representing the conservation of charge. One can imagine also an
asymmetrical distribution of surface charges with a domain wall cutting the
circle of the electron pool as a segment and rotating as a whole in the
direction determined by $\mathbf{B}$. By now, the experiment on the resonant
photovoltaic effect~\cite{KonCheKon-2012} favors the symmetrical distribution
indicated in Fig.~\ref{f20}.

If the domain wall is sharp, the separation of charges shown in Fig.~\ref{f20} is
stable while the resonant MW excitation keeps $\sigma _{xx}=0$ in the
domains. Assume some electrons by chance have moved from the area $S_{%
\mathrm{L}}$ to the area $S_{\mathrm{H}}$. This will make $\sigma _{xx}>0$
in both areas, and the same amount of electrons will return back due to the
inner electric field induced by charge separation. In the contrary case,
when some electrons have moved from the area $S_{\mathrm{H}}$ to the area $%
S_{\mathrm{L}}$, in both domains $\sigma _{xx}$ becomes negative, and the
same amount of electrons will return back moving uphill. Thus, under the
condition of Eq.~(\ref{e40}) the domain structure is in a dynamic
equilibrium. The same arguments are valid also for an asymmetrical
distribution of surface charges discussed above.

Formation of the domain structure caused by negative conductivity effects
explains electron redistribution observed under resonant MW
radiation~\cite{KonCheKon-2012}. According to theoretical estimations the displaced
fraction of electrons can be of the order of $N_{e}$ which agrees with
observations. The domain structure of electron density can explain also
self-generated oscillations observed in the experiment~\cite{KonWatKon-2013}.
If the inter-edge profile of the density domain wall is sufficiently smooth,
a negative dissipative conductivity $\sigma _{xx}$ can be ascribed to a
narrow strip of the domain wall. It is quite obvious that this can lead to a
negative damping and to self-generation of the inter-edge modes (the same
argument is applicable to the edge excitations propagating along the edge of
the electron pool if the net conductivity within the edge profile is
negative).

There are two kinds of inter-edge modes: inter-edge
magnetoplasmons (IEMP)~\cite{MikVol-1992}, and boundary displacement waves
(BDW)~\cite{Mon-1995,Mon-2001-book}. Both of them have similar gapless
spectrums at strong magnetic fields, and usually they are strongly coupled.
For example, the frequency of IEMP decreases with $B$ as~\cite{MikVol-1992}%
\begin{equation}
\omega _{\mathrm{IEMP}}=2q_{y}\left( \sigma _{yx}^{\mathrm{H}}-\sigma _{yx}^{%
\mathrm{L}}\right) \left( \ln \frac{1}{\left\vert q_{y}\right\vert w}%
+C\right) ,  \label{e41}
\end{equation}%
where $\sigma _{yx}^{\mathrm{H/L}}\propto n_{\mathrm{H/L}}/B$, the constant $%
C$ depends on details of the density profile, $w$ is the width of the
transition layer, and $q_{y}$ is the wavevector component along the
boundary. The frequency $\omega _{\mathrm{BDW}}$ of the BDW differs from Eq.~(%
\ref{e41}) only by a numerical factor of the order of unity~\cite{Mon-1995}.
Experimental observations of the IEMP~\cite{SomSteHei-1995} and
BDW~\cite{KirSomWae-1995,YamAraYay-2012,YamAraYay-2013} are in agreement with the theories. It
should be noted that the BDW can propagate in an incompressible 2D electron
liquid. For the typical length of the domain wall, the both frequencies $%
\omega _{\mathrm{IEMP}}$ and $\omega _{\mathrm{BDW}}$ belong to the
audio-frequency range which is approximately the same as that reported for
self-generated oscillations observed~\cite{KonWatKon-2013}. Thus,
experimental observation of self-generated audio-frequency oscillations can
be considered as a convincing evidence for negative conductivity coexisting
with the ZRS of surrounding domains.

Here we would like to note an interesting analogy with current domains in
a 2D electron gas formed in semiconductors. Nonuniformities in the electron density
can lead, under certain circumstances, to time-dependent domain patterns in
the microwave-induced ZRS~\cite{FinHal-2009}. This result
was used for explanation of random telegraph signals observed in the
zero-resistance regime~\cite{DorPfeSme-2011}. As noted above, electrons
on liquid helium form a highly correlated Coulomb liquid which in some respects
resembles the Wigner solid. Therefore, electron velocities at the both sides of the
domain wall are expected to be the same, and the results obtained for semiconductor
electrons cannot be applied directly to the SE system.

The appearance of two critical electron densities $n_{\mathrm{H}}$ and $n_{%
\mathrm{L}}$ restricting the instability range provides also an insight into
the nature of the incompressible state observed for the system of two
coupled 2D electron gases~\cite{CheWatKon-2015}. One can expect anomalies on
the dependence $n_{\mathrm{eM}}\left( V_{\mathrm{g}}\right) $ near $n_{%
\mathrm{H}}$ and $n_{\mathrm{L}}$. In Ref.~\cite{Mon-2016}, the plateau value
$n_{\mathrm{c}}$ of Fig.~\ref{f18} was naturally ascribed to $n_{\mathrm{H}}$%
, while the density to which the solid line falls down at $V_{g}\simeq 5.13%
\,\mathrm{V}$ was ascribed to $n_{\mathrm{L}}$ (note that the later density
point is also practically independent of $N_{e}$ according to the
data~\cite{CheWatKon-2015}). This assumption is in accordance with the conclusion that
the system is unstable in the region $n_{\mathrm{L}}<n_{e}<n_{\mathrm{H}}$.
Thus, we have $n_{\mathrm{H}}\simeq 3.4\cdot 10^{6}\,\mathrm{cm}^{-2}$ and $n_{%
\mathrm{L}}\simeq 1.2\cdot 10^{6}\,\mathrm{cm}^{-2}$. It is remarkable that
the small upper plateau of $n_{\mathrm{eM}}\left( V_{\mathrm{g}}\right) $
and the plateau of $n_{\mathrm{gM}}\left( V_{\mathrm{g}}\right) $ formed at $%
V_{G}\approx 4\,\mathrm{V}$ resulting in the second critical density $n_{%
\mathrm{gM}}=n_{\mathrm{c}}^{\left( 2\right) }\simeq 1\cdot 10^{6}\,\mathrm{cm}%
^{-2}$ are close to the condition $n_{\mathrm{c}}^{\left( 2\right) }=n_{%
\mathrm{L}}$.

Of course, the experimental conditions of Ref.~\cite{CheWatKon-2015} were
different from usual conditions of MICO
experiments~\cite{KonCheKon-2012,KonWatKon-2013} because
in this work density domains were created artificially by applying different
potentials to the guard ($V_{\mathrm{g}}$) and central ($V_{\mathrm{B}}$)
electrodes of Corbino geometry. In this case, the presence of a neighbor 2D
electron gas located in the range of the guard electrodes affects properties
of inner electrons, and compressibility $\chi $ defined in Eq.~(\ref{e37})
using $n_{\mathrm{e}}\left( V_{\mathrm{g}}\right) $ cannot be ascribed
solely to electrons of the central region. Nevertheless, $\chi $ measured
gives important information about 2D electron gases under MW irradiation.

The regime of the experiment~\cite{CheWatKon-2015} can be related to the
domain wall structure discussing here only if the both densities $n_{\mathrm{%
e}}$ and $n_{\mathrm{g}}$ belongs to the unstable region $n_{\mathrm{L}}<n_{%
\mathrm{g}},n_{\mathrm{e}}<n_{\mathrm{H}}$. Luckily, such a regime occurs
when $n_{\mathrm{e}}$ is a bit lower than $n_{\mathrm{c}}$ ascribed to $n_{%
\mathrm{H}}$. As noted in Ref.~\cite{Mon-2016}, in this case, $n_{\mathrm{g}}$
is somewhat larger than $n_{\mathrm{L}}$, which allows explaining the
stability of the state $n_{\mathrm{e}}=n_{\mathrm{c}}=n_{\mathrm{H}}$ for
finite ranges of $V_{\mathrm{g}}$ and $N_{e}$. According to the theoretical analysis of
the experimental situation, the whole electron system should be separated in
two parts with fixed densities $n_{\mathrm{H}}$ and $n_{\mathrm{L}}$ with a
domain wall positioned between the edge of the center electrode ($R_{I}=0.7%
\,\mathrm{cm}$) and the edge of the guard electrode ($R_{\mathrm{G}}=1.3%
\,\mathrm{cm}$). The high density domain with $n_{e}=n_{\mathrm{H}}$
is placed in the center of the
electron pool because the potential applied to the guard electrode is opposite
to the usual case (SE are attracted to the guard region).
For such a distribution of SE, a change in the total number
of electrons $N_{e}$ can only shift the position of the domain wall $R_{%
\mathrm{D}}$ inside the region $R_{I}<R_{\mathrm{D}}<R_{\mathrm{G}}$ which
cannot affect the number of electrons coupled to the central electrode and
the current $I$. This explains independence of the number of electrons coupled to
central electrode of the total number $N_{e}$ when $n_{\mathrm{e}}=n_{\mathrm{c}}=n_{\mathrm{H}}$.

One can imagine a distribution of SE with a domain wall placed inside the
central electrode $R_{\mathrm{D}}<$ $R_{I}$. In this case, the average
density of electrons coupled to the central electrode will be dependent on $%
N_{e}$, while the the average density of electrons coupled to the guard
electrode will be approximately constant. This remarkably reminds the
experimental situation related to the second plateau observed in the region II of Fig.~\ref{f18}
(at $V_{\mathrm{g}}\sim 4\,\mathrm{V}$).

\section{Polarization dependence}
\label{sec:8}

For the inter-subband mechanism of MICO and ZRS, the crucial point is that
the MW field $\mathbf{E}_{\mathrm{mw}}$ should have a vertical component $E_{%
\mathrm{mw}}^{\left( z\right) }\neq 0$. The in-plane component of this field
is unimportant because at typical MW powers used in the experiments with
inter-subband excitation~\cite{KonKon-2009,KonKon-2010} the photon-assisted
scattering of SE can be neglected due to $m_{e}\gg m_{e}^{\ast }$. At the
same time, for the theory of photogalvanic current~\cite{EntMag-2013}, the
presence of the in-plane component of $\mathbf{E}_{\mathrm{mw}}$ is crucial,
and its contribution to the photovoltaic effect can be verified by using the
MW field of pure vertical polarization $E_{\mathrm{mw}}^{\left( z\right)
}=E_{\mathrm{mw}}$.

Regarding the intra-subband mechanisms of MIRO and MICO, the polarization
dependence of oscillations was considered as a major experimental test for
theoretical models. Among these mechanisms, there is a large group of models
whose description is based on the theory of photon-assisted scattering off
disorder. The displacement and inelastic mechanisms discussed in
Ref.~\cite{DmiMirPol-2012} and in Subsection~\ref{subsec:2-1}
(for SE on liquid helium) represent the most
elaborated examples of such models. Both the displacement and inelastic
models give satisfactory descriptions of MIRO in semiconductor
heterostructures, if the dependence of MIRO on the direction of circular
polarization is not considered. The results of theories based on the
photon-assisted scattering are very sensitive to the direction of circular
polarization. At the same time, the MIRO observed in semiconductor
heterostructures are notably immune to the sense of circular
polarization~\cite{SmeGorKli-2005} or have a very weak dependence on the direction of circular
polarizations in the terahertz range~\cite{HerDmiGan-2016} which is at odds
with existing theories of MIRO. Therefore, experiments on MICO caused by
intra-subband excitation of SE on liquid helium were expected to help with
understanding of the origin of these oscillations and ZRS.

It should be noted firstly, that the displacement model gives a strong
dependence of MICO on the direction of linear MW polarization shown in Fig.%
~\ref{f3}. In contrast, the MICO obtained in the inelastic model have the
linear polarization immunity: $\bar{\chi}_{\bot }=\bar{\chi}_{\Vert }$ as
indicated in Eq.~(\ref{e22}). This immunity appears because the transition
rate $r_{n,n^{\prime }}$ given in Eq.~(\ref{e20}) is finite in the limit $%
E_{\Vert }\rightarrow 0$, and integration over the angle of the vector $%
\mathbf{q}$ results in equal averaging of $q_{x}^{2}$ and $q_{y}^{2}$
entering the polarization parameter $\beta _{p,\mathbf{q}}$ of Eq.~(\ref{e10}%
). Obviously, the difference of the results given by these two models cannot
be verified in the usual experimental setup employing the Corbino disks. On
the other hand, the Corbino geometry of the experimental cell is very
suitable for the test on the circular-polarization dependence of MICO.

Consider circular polarizations of the in-plane MW field ($p=+$ and $p=-$)
determined by the following values of parameters $a$ and $b$ entering the
definition of $\beta _{p,\mathbf{q}}$: $a_{\pm }=1$, $b_{\pm }=\pm 1$. The
probability of one-photon-assisted scattering is proportional to $%
J_{1}^{2}\left( \beta _{p,\mathbf{q}}\right) \simeq \beta _{p,\mathbf{q}%
}^{2}/4$, if the parameter $\beta _{p,\mathbf{q}}$ is small. Therefore, the
ratio of MW-induced corrections to the dc dissipative conductivity obtained
for different directions of circular polarization is described as%
\begin{equation}
\frac{\Delta \sigma _{xx}^{\left( +\right) }}{\Delta \sigma _{xx}^{\left(
-\right) }}=\frac{\left( \omega /\omega _{c}+1\right) ^{2}}{\left( \omega
/\omega _{c}-1\right) ^{2}}  \label{e42}
\end{equation}%
This equation is valid for the both displacement and inelastic mechanisms of
MICO, because, for a circular polarization, $\beta _{p,\mathbf{q}}$ is
independent of the direction of the wave vector $\mathbf{q}$. The ratio of
Eq.~(\ref{e42}) is large for $m=2$ (it equals $9$) and $m=3$ (it equals $4$),
but it approaches unity if $m$ increases.

\begin{figure}
 \center{ \includegraphics[width=9cm]{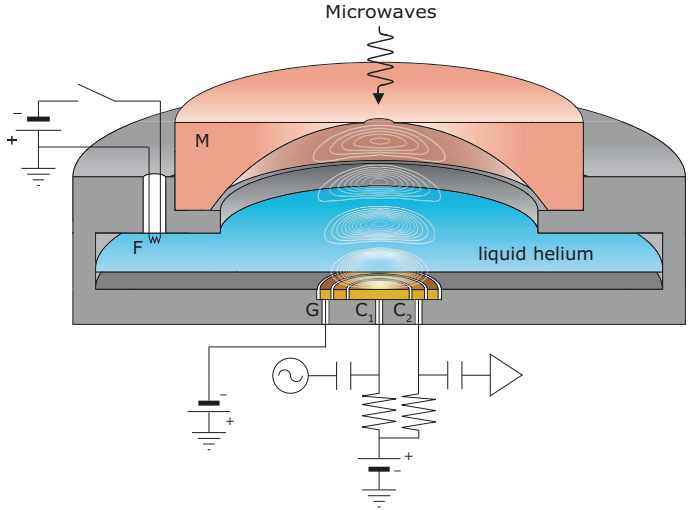}}
\caption{(Color online) Sketch of the experimental cell. White
lines show electric field contours calculated for $\mathrm{TEM}_{003}$ mode.
The pool of 2D electrons (not shown) is formed on the
surface of liquid helium above electrodes $C_1$ (central) and $C_2$ (middle)~\cite{YamMonKon-2015}
 }
\label{f21}
\end{figure}

Circular-polarization-dependent studies of MICO~\cite{ZadMonKon-2018} were
done in a 2D electron system formed on the free surface of liquid $^{3}%
\mathrm{He}$, which was contained in a closed cylindrical cell and cooled to
$T=0.2\,\mathrm{K}$. The magnetic field $B$ was applied perpendicular to the
liquid surface, and the longitudinal conductivity of electrons $\sigma _{xx}$
was measured by the capacitive-coupling method using a pair of gold-plated
concentric circular electrodes (Corbino disk) placed beneath and parallel to
the liquid surface. Conductivity oscillations were excited by the electric
field component of the fundamental $\mathrm{TEM}_{002}$ mode in a
semiconfocal Fabry-Perot resonator~\cite{Kog-1966}. The resonator is formed
by the Corbino disk acting as a flat reflecting mirror and a copper concave
mirror placed above and parallel to the Corbino disk (similar to the cell of the experiment~\cite{YamMonKon-2015} shown in
Fig.~\ref{f21}). At liquid helium temperatures, the $\mathrm{TEM}_{002}$ mode had the
frequency $\omega /2\pi \approx 35.21\,\mathrm{GHz}$, and the quality
factor was about $10^{4}$.

\begin{figure}
 \center{ \includegraphics[width=10cm]{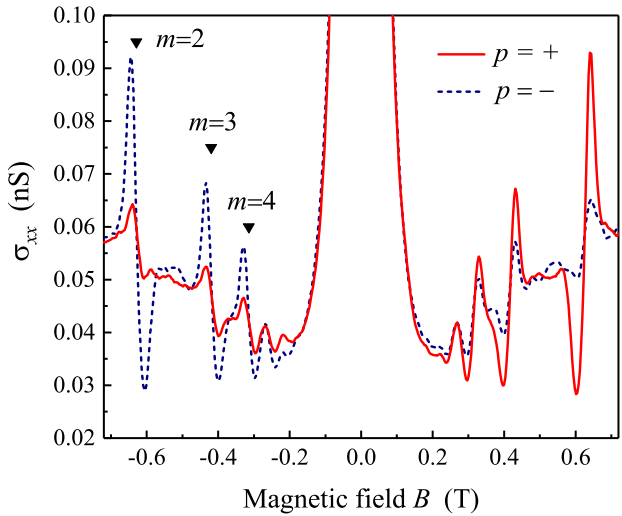}}
\caption{(Color online) $\sigma _{xx}$ of SE at $T=0.2\,\mathrm{K}$ and $n_{e}=5.1\cdot 10^{6}\,\mathrm{cm}^{-2}$ vs. $B$ for two directions of circular polarization of the MW field ($\omega /2\pi =35.213\,\mathrm{GHz}$): $p=+ $ (red solid) and $p=-$ (blue dashed).
Black triangles indicate the values of $B$ such that $\omega /\omega _c=m$~\cite{ZadMonKon-2018}
 }
\label{f22}
\end{figure}

The $\sigma _{xx}$ data~\cite{ZadMonKon-2018} plotted as a function of the
magnetic field $B$ are shown in Fig.~\ref{f22} for two opposite directions of
circular polarization $p=\pm $. The strong dependence of the amplitude of
oscillations on the direction of circular polarization indicated in this
figure is at least in qualitative agreement with predictions of the theories
based on photon-assisted scattering. In the region $B>0$, the MICO amplitude
is larger for $p=+$ (red solid curve), while in the opposite region $B<0$, it is
larger for $p=-$ (blue dashed curve).

\begin{figure}
 \center{ \includegraphics[width=9cm]{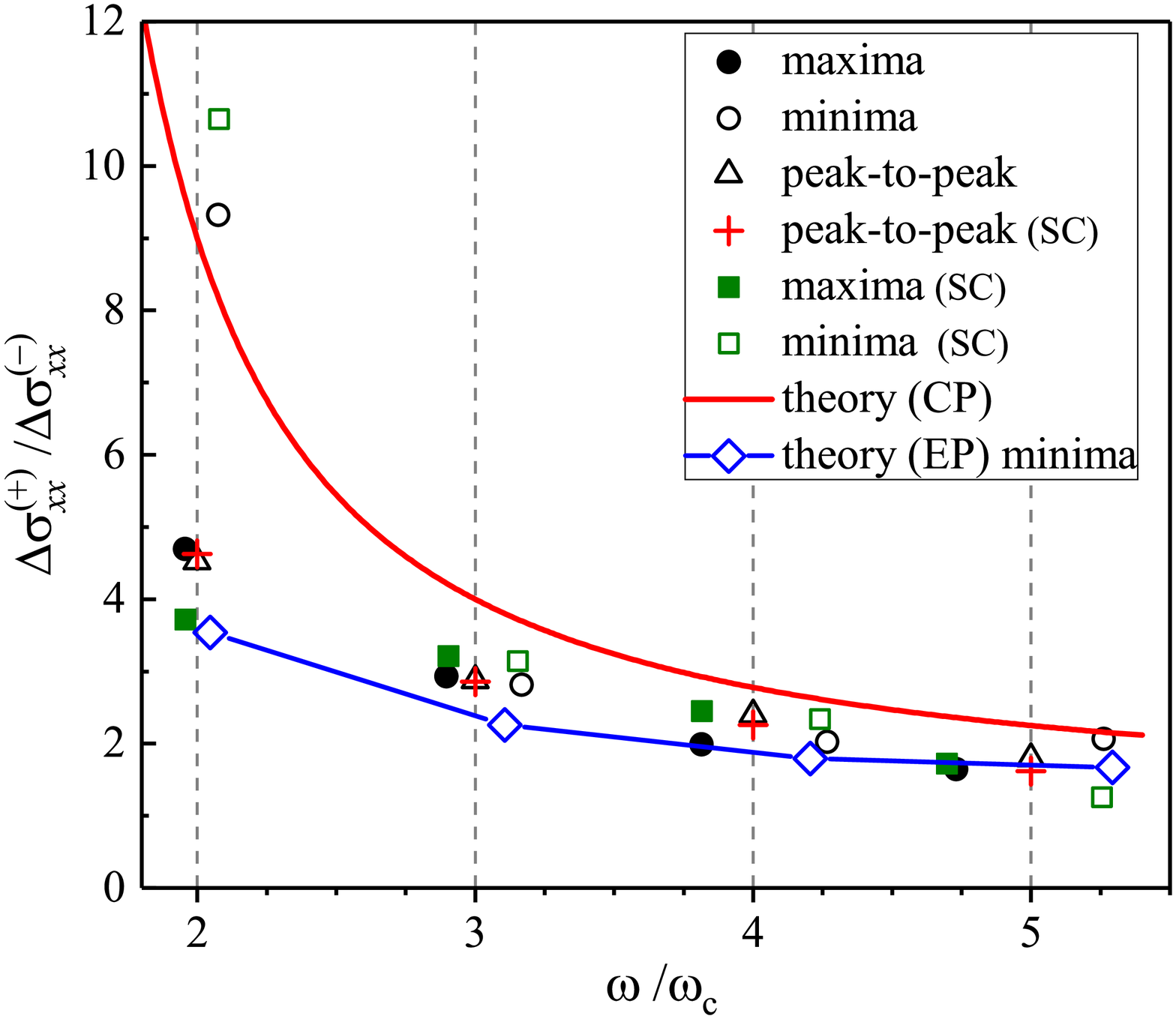}}
\caption{(Color online) The ratio $\Delta \sigma _{xx}^{\left( +\right) }/\Delta \sigma _{xx}^{\left(
-\right) }$ versus $\omega / \omega _{c}$: results obtained from experimental data for conductivity maxima (filled circles and squares) and
minima (open circles and squares), the ratio of peak-to-peak amplitudes (triangles and crosses), the ratio of amplitudes obtained for positive and negative $B$ using the red solid curve (SC) of Fig.~\ref{f22},
theory [Eq.~(\ref{e3})] calculated for the circular polarization (CP, red solid line),
and theory using the elliptic polarization (EP) with $a_{\pm }=0.7$ and $b_{\pm }=\pm 1.3$
calculated for minima (open rhombuses)~\cite{ZadMonKon-2018}
 }
\label{f23}
\end{figure}

The quantitative comparison of the theory and experiment is shown in Fig.~\ref{f23}.
Here, the red solid line represents the ratio $\Delta \sigma
_{xx}^{\left( +\right) }/\Delta \sigma _{xx}^{\left( -\right) }$ given in
Eq.~(\ref{e42}). Experimental results extracted from the data of Fig.~\ref{f22}
were plotted separately for maxima (solid circles and squares) and
minima (open circles and squares), because the ratio $\Delta \sigma
_{xx}^{\left( +\right) }/\Delta \sigma _{xx}^{\left( -\right) }$ depends
strongly on $B$. The accuracy of this procedure was confirmed by the ratio
of peak-to-peak amplitudes (triangles) which does not depend on the
background. To ensure that a possible power difference did not affect the
results, the ratio of respective amplitudes obtained at positive and
negative $B$ from the same curve (red solid) of Fig.~\ref{f22} was plotted as squares and
crosses. The experimental data show that $\Delta \sigma _{xx}^{\left(
+\right) }/\Delta \sigma _{xx}^{\left( -\right) }$ increases with lowering $%
\omega /\omega _{c}$ in accordance with the theory. Still, at average,
they are lower than theoretical values obtained for pure circular
polarizations by about $1.4$. This numerical discrepancy was
explained~\cite{ZadMonKon-2018} by deviation from circularity (ellipticity) of the MW
field. Note that even a relatively small response at CR conditions for the MW field
with $p=-$, which was observed in the experiment~\cite{ZadMonKon-2018},
suggests substantial deviations of parameters $a_{\pm }$ and $%
\left\vert b_{\pm }\right\vert $ from unity. In Fig.~\ref{f23}, the open-rhombus
symbols show the result of calculations performed for polarization
parameters $a_{\pm }$ and $b_{\pm }$ estimated from the photocurrent
response at CR conditions. Thus, the dependence of MICO on the direction of
circular polarization of the MW observed~\cite{ZadMonKon-2018} is in a good
(even numerical) agreement with the theory based on photon-assisted
scattering.

Regarding the mysterious immunity of MIROs to the sense of circular
polarization reported previously~\cite{SmeGorKli-2005} for the 2D electron gas
in GaAs/AlGaAs heterostructures, it could be a property of particular
semiconductor samples. Recent numerical simulations~\cite{CheShe-2018} of
electron dynamics in a vicinity of impurity indicate that the MW irradiation
generates a rotating charge density vortex whose field eventually leads to
immunity of MIRO to the sense of circular polarization at high electron
densities typical for semiconductor systems. The densities of SE on liquid
helium usually are several orders of magnitude smaller than in semiconductor
heterostructures.

\begin{acknowledgements}
The work of D. K. is supported by an internal grant from Okinawa Institute of Science and Technology (OIST)
Graduate University. We are grateful to Kimitoshi Kono, Alexei Chepelianskii, Masamitsu Watanabe,
and Konstantin Nasedkin for fruitful collaborations on the topics included in this review.
\end{acknowledgements}


\begin{thebibliography}{}
%
%
\bibitem{ColCoh-1969}  M.W. Cole and M.H. Cohen, Phys. Rev. Lett. \textbf{23},
1238 (1969)

\bibitem{Shi-1970}  V.B. Shikin, Soviet Phys. JETP \textbf{31}, 936 (1970)
[Zh. Eksp. Teor. Fiz. \textbf{58}, 1748 (1970)]

\bibitem{LeiWan-1979} P. Leiderer and M. Wanner, Phys. Letters A, \textbf{73}, 189 (1979)

\bibitem{Ede-1980}  V.S. Edel'man, Sov. Phys. Usp. \textbf{23}, 227 (1980)
[Usp. Fiz. Nauk \textbf{130}, 675 (1980)]

\bibitem{EtzGomLei-1984}  H. Etz, W. Gombert, W. Idstein and P.
Leiderer, Phys. Rev. Lett. \textbf{53}, 2567 (1984)

\bibitem{AndFowSte-1982}  T. Ando, A.B. Fowler, and F. Stern, Rev. Mod. Phys.
\textbf{54}, 437 (1982)

\bibitem{KliDorPep-1980} von K. Klitzing, G. Dorda, and M. Pepper,
Phys. Rev. Lett. \textbf{45}, 494 (1980)

\bibitem{TsuStoGos-82} D.C. Tsui, H.L. Stormer, and A.C.
Gossard, Phys. Rev. Lett. \textbf{48}, 1559 (1982)

\bibitem{GriAda-1979}  C.C. Grimes and G. Adams: Phys. Rev. Lett. \textbf{42},
795 (1979)

\bibitem{MonSyv-2012} Yu.P. Monarkha and V.E. Syvokon, Low Temp. Phys. \textbf{38}, 1067 (2012)
[Fiz. Nizk. Temp. \textbf{38}, 1355 (2012)] 

\bibitem{And-1997} E. Y. Andrei, {\it Electrons on Helium and Other Cryogenic
Substrates}, Kluwer Academic, Dordrecht (1997)

\bibitem{MonTesWyd-2002} Yu. P. Monarkha, E. Teske, and P. Wyder, Phys. Rep. \textbf{370}, 1 (2002)

\bibitem{MonKon-book} Yu.P. Monarkha and K. Kono, {\it Two-Dimensional Coulomb Liquids
and Solids}, Springer-Verlag, Berlin (2004)

\bibitem{ZudDuRen-2001} M.A. Zudov, R.R. Du, J.A. Simmons, and J.L. Reno, Phys. Rev. B \textbf{64}, 201311 (2001)

\bibitem{YeEngRen-2001} P.D. Ye, L.W. Engel, D.C. Tsui, J.A. Simmons, J.R. Wendt, G.A. Vawter, and J.L. Reno, Appl. Phys. Lett. 79,
2193 (2001)

\bibitem{ManSmeKli-2002} R. Mani, J. H. Smet, K. von Klitzing, V. Narayanamurti, W. B. Johnson,
and V. Umansky, Nature (London) \textbf{420}, 646 (2002)

\bibitem{ZudDuWes-2003} M. A. Zudov, R. R. Du, L. N. Pfeiffer, and K.W. West,
Phys. Rev. Lett. \textbf{90}, 046807 (2003)

\bibitem{AndAleMil-2003} A.V. Andreev, I.L. Aleiner, and A.J. Millis, Phys. Rev. Lett. \textbf{91}, 056803 (2003)

\bibitem{Ryz-1969} V.I. Ryzhii, Sov. Phys. Solid State \textbf{11}, 2078 (1970)
[Fiz. Tverd. Tela (Leningrad) \textbf{11}, 2577 (1969)]

\bibitem{DurSachRea-2003} A.C. Durst, S. Sachdev, N. Read, and S.M. Girvin, Phys. Rev.
Lett. \textbf{91}, 086803 (2003)

\bibitem{RyzChaSur-2004} V. Ryzhii, A. Chaplik, and R. Suris, JETP Letters \textbf{80},
363 (2004) [Pis'ma v ZhETF, \textbf{80}, 412 (2004)] 

\bibitem{DmiMirPol-2003} I.A. Dmitriev, A.D. Mirlin, and D.G. Polyakov,
Phys. Rev. Lett. \textbf{91}, 226802 (2003)

\bibitem{DmiVavAle-2005} I.A. Dmitriev, M.G. Vavilov, I.L. Aleiner, A.D.
Mirlin, and D.G. Polyakov, Phys. Rev. B \textbf{71}, 115316 (2005)

\bibitem{DmiMirPol-2012} I.A. Dmitriev, A.D. Mirlin, D.G. Polyakov, and M.A.
Zudov, Rev. Mod. Phys. \textbf{84}, 1709 (2012)

\bibitem{ZudMirEbn-2014} M.A. Zudov, O.A. Mironov, Q.A. Ebner, P.D. Martin, Q.
Shi, and D.R. Leadley, Phys. Rev. B \textbf{89}, 125401 (2014)

\bibitem{KarShcSme-2016} D.F. Karcher, A.V. Shchepetilnikov, Yu.A. Nefyodov, J. Falson, I.A. Dmitriev,
Y. Kozuka, D. Maryenko, A. Tsukazaki, S.I. Dorozhkin, I.V. Kukushkin, M. Kawasaki,
and J.H. Smet, Phys. Rev. B \textbf{93}, 041410(R) (2016). 

\bibitem{KonKon-2009} D. Konstantinov and K. Kono, Phys. Rev. Lett. \textbf{103},
266808 (2009). 

\bibitem{KonKon-2010} D. Konstantinov and K. Kono, Phys. Rev. Lett. \textbf{105},
226801 (2010). 

\bibitem{Mon-2011} Yu.P. Monarkha, Low Temp. Phys. \textbf{37}, 90 (2011)
[Fiz. Nizk. Temp. \textbf{37}, 108 (2011)] 
Yu.P. Monarkha, Low Temp. Phys. \textbf{37}, 655 (2011)
[Fiz. Nizk. Temp. \textbf{37}, 829 (2011)]  

\bibitem{Mon-2012} Yu.P. Monarkha, Low Temp. Phys. \textbf{38}, 451 (2012)
[Fiz. Nizk. Temp. \textbf{38}, 579 (2012)]

\bibitem{KonCheKon-2012} D. Konstantinov, A. Chepelianskii, and K. Kono, J. Phys. Soc.
Jpn., \textbf{81}, 093601 (2012)

\bibitem{KonWatKon-2013} D. Konstantinov, M. Watanabe, and K. Kono, J. Phys. Soc.
Jpn., \textbf{82} 075002 (2013)

\bibitem{CheWatKon-2015} A.D. Chepelianskii, M. Watanabe, K. Nasyedkin, K. Kono, and D. Konstantinov,
Nature Communications, \textbf{6}, 7210 (2015)

\bibitem{Mon-2016} Yu.P. Monarkha, Low Temp. Phys. \textbf{42}, 441 (2016)
[Fiz. Nizk. Temp. \textbf{42}, 567 (2016)] 

\bibitem{Mon-2014} Yu.P. Monarkha, Low Temp. Phys. \textbf{40}, 482 (2014)
[Fiz. Nizk. Temp. \textbf{40}, 623 (2014)] 

\bibitem{YamMonKon-2015} R. Yamashiro, L.V. Abdurakhimov, A.O. Badrutdinov, Yu.P. Monarkha,
and D. Konstantinov, Phys. Rev. Lett. \textbf{115}, 256802 (2015)

\bibitem{ZadMonKon-2018} A.A. Zadorozhko, Yu.P. Monarkha, and D. Konstantinov,
Phys. Rev. Lett. \textbf{120}, 046802 (2018) 

\bibitem{VavAle-2004} M.G. Vavilov and I.L. Aleiner, Phys. Rev. B \textbf{69}, 035303 (2004)

\bibitem{Mon-2017} Yu.P. Monarkha, Low Temp. Phys. \textbf{43}, 650 (2017)
[Fiz. Nizk. Temp. \textbf{43}, 819 (2017)] 

\bibitem{ShiMon-74}  V.B. Shikin, Yu.P. Monarkha: J. Low Temp. Phys.
\textbf{16}, 193 (1974)

\bibitem{Par-2004} K. Park, Phys. Rev. B \textbf{69}, 201301(R) (2004)

\bibitem{Hus-1953} K. Husimi, Prog. Theor. Phys. \textbf{9}, 381 (1953)

\bibitem{AndUem-1974} T. Ando and Y. Uemura: J. Phys. Soc. Jpn. \textbf{36},
959 (1974)

\bibitem{Ger-1976}  R.R. Gerhardts: Surf. Sci. \textbf{58}, 227 (1976)

\bibitem{And-1974}  T. Ando: J. Phys. Soc. Jpn. \textbf{37}, 622 (1974)

\bibitem{FanDykLea-1997}  C. Fang-Yen, M.I. Dykman, and M.J. Lea, Phys. Rev. B
\textbf{55}, 16272 (1997) 

\bibitem{MonSokStu-2010} Yu.P. Monarkha, S.S. Sokolov, A.V. Smorodin, and N.
Studart, Low Temp. Phys. \textbf{36}, 565 (2010) [Fiz. Nizk. Temp. \textbf{36}, 711 (2010)]

\bibitem{Mon-2013_LTP} Yu.P. Monarkha, Low Temp. Phys. \textbf{39}, 828 (2013)
[Fiz. Nizk. Temp. \textbf{39}, 1068 (2013)] 

\bibitem{Mon-2013_JETP} Yu.P. Monarkha, JETP Letters, \textbf{98}, 9 (2013)
[Pisma v ZhETF, \textbf{98}, 12 (2013)] 

\bibitem{ZipBroGri-1976} C.L. Zipfel, T.R. Brown, and C.C. Grimes,
Phys. Rev. Lett. \textbf{37}, 1760 (1976)

\bibitem{KonMonKon-2007} D. Konstantinov, H. Isshiki, Yu. Monarkha, H. Akimoto, K. Shirahama, and K. Kono,
Phys. Rev. Lett. \textbf{98}, 235302 (2007) 

\bibitem{DykKha-1979} M.I. Dykman and L.S. Khazan, Sov. Phys. JETP \textbf{50}, 747 (1979)
[Zh. Eksp. Teor. Fiz. \textbf{77}, 1488 (1979)]

\bibitem{KonMonKon-2013} D. Konstantinov, Yu. Monarkha, and K. Kono, Phys. Rev. Lett.
\textbf{111}, 266802 (2013) 

\bibitem{Zud-2004} M. A. Zudov, Phys. Rev. B \textbf{69}, 041304(R) (2004)

\bibitem{WilPheWes-2004} R.L. Willett, L.N. Pfeiffer, and K.W. West, Phys. Rev. Lett.
\textbf{93}, 026804 (2004) 

\bibitem{EntMag-2013} M.V. Entin and L.I. Magarill, JETP Letters \textbf{98}, 816 (2013)
[Pisma v ZhETF \textbf{98}, 919 (2013)] 

\bibitem{CraWil-1972} R.S. Crandall, R. Williams. Phys. Rev. A \textbf{5}, 2183 (1972)

\bibitem{MikVol-1992} S.A. Mikhailov, V.A. Volkov, J. Phys.: Condens. Matter \textbf{4},
6523 (1992). 

\bibitem{Mon-1995} Yu.P. Monarkha, Low Temp. Phys. \textbf{21}, 458 (1995)
[Fiz. Nizk. Temp. \textbf{21}, 589 (2013)]

\bibitem{Mon-2001-book} Yu.P. Monarkha, in \textit{Edge Excitations of Low-Dimensional
Charged Systems}, ed. by O. Kirichek, Nova Science Publishers Inc.,
New York, Chapt. 2, pp. 4974 (2001)

\bibitem{SomSteHei-1995} P.K.H. Sommerfeld, P.P. Steijaert, P.J.M. Peters, and
R.W. van der Heijden, Phys. Rev. Lett. \textbf{74}, 2559 (1995)

\bibitem{KirSomWae-1995} O.I. Kirichek, P.K.H. Sommerfeld, Yu.P. Monarkha, P.J.M. Peters, Yu.Z. Kovdrya,
P.P. Steijaert, R.W. van der Heijden, and A.T.A.M. de Waele, Phys. Rev. Lett. \textbf{74}, 1190 (1995)

\bibitem{YamAraYay-2012} S. Yamanaka, T. Arai, A. Sawada, A. Fukuda, and H. Yayama, EPL, \textbf{100} 17009 (2012)

\bibitem{YamAraYay-2013} S. Yamanaka, T. Arai, A. Sawada, A. Fukuda, and H. Yayama, Low Temp. Phys. \textbf{39}, 842 (2013)
[Fiz. Nizk. Temp. \textbf{39}, 1086 (2013)] 

\bibitem{FinHal-2009} I.G. Finkler and B.I. Halperin, Phys. Rev. B \textbf{79}, 085315 (2009)

\bibitem{DorPfeSme-2011} S.I. Dorozhkin, L. Pfeiffer, K. West, K. von Klitzing and J.H. Smet,
Nature Physics \textbf{7}, 336 (2011)

\bibitem{SmeGorKli-2005} J.H. Smet, B. Gorshunov, C. Jiang, L. Pfeiffer, K. West,
V. Umansky, M. Dressel, R. Meisels, F. Kuchar, and
K. von Klitzing, Phys. Rev. Lett. \textbf{95}, 116804 (2005)

\bibitem{HerDmiGan-2016} T. Herrmann, I.A. Dmitriev, D.A. Kozlov, M. Schneider,
B. Jentzsch, Z.D. Kvon, P. Olbrich, V.V. Belkov, A. Bayer,
D. Schuh, D. Bougeard, T. Kuczmik, M. Oltscher, D. Weiss,
and S.D. Ganichev, Phys. Rev. B \textbf{94}, 081301(R) (2016)

\bibitem{Kog-1966} H. Kogelnik and T. Li, Appl. Opt. \textbf{5}, 1550 (1966)

\bibitem{CheShe-2018} A.D. Chepelianskii and D.L. Shepelyansky, Phys. Rev. B \textbf{97}, 125415 (2018)


\end{thebibliography}
\end{document}